\documentstyle[a4]{elsartwb}

\input epsf

\begin{document}

\begin{frontmatter}

\title{Solitons in nonlocal chiral quark models\thanksref{grants}}
\thanks[grants]{Research supported in part by the
Scientific and Technological Cooperation Joint Project between
Poland and Slovenia, financed by the
Ministry of Science of Slovenia and the Polish State Committee for
Scientific Research, and by the Polish State Committee for
Scientific Research, grant 2 P03 09419}
\thanks[emails]{%
\hspace{0mm} E-mail: broniows@solaris.ifj.edu.pl, 
bojan.golli@ijs.si, ripka@cea.fr}
\author[INP]{Wojciech Broniowski}, 
\author[LJ]{Bojan Golli}, 
and
\author[Saclay,Trento]{Georges Ripka}
\address[INP]{The H. Niewodnicza\'nski Institute of Nuclear
Physics, ul. Radzikowskiego 152, PL-31342 Krak\'ow, Poland}
\address[LJ]
{Faculty of Education, University of Ljubljana and J.~Stefan Institute,
Jamova 39, P.O.~Box 3000, 1001 Ljubljana, Slovenia}
\address[Saclay] {Service de Physique Th\'eorique, Centre d'Etudes de Saclay,
F-91191 Gif-sur-Yvette Cedex, France}
\address[Trento]{ECT${}^\ast$, Villa Tombosi, 
I-38050 Villazano (Trento), Italy}

\begin{abstract}
Properties of hedgehog solitons in a chiral quark model with nonlocal
regulators are described. We discuss the formation of the hedgehog
soliton, the quantization of the baryon number,
the energetic stability, the gauging and construction of Noether currents 
with help of 
path-ordered $P$-exponents, and the evaluation of observables. 
The issue of nonlocality is thoroughly discussed, with a focus on
contributions to observables related to the Noether currents.
It is shown that with typical model 
parameters the solitons are not far from the 
weak nonlocality limit. The methods developed are applicable to
solitons in models with separable nonlocal four-fermion interactions.
\end{abstract}


\end{frontmatter}


\section{Introduction}

\label{sec:intro} Effective chiral quark models have been used extensively
to describe the low-energy phenomena associated with the dynamical breaking
of the chiral symmetry. Of particular interest are models which include the
Dirac sea of quarks, since they allow for a common description of mesons and
baryons. The models share the feature of having an attractive four-quark
interaction which is either:

\begin{itemize}
\item  postulated \cite{njl};

\item  derived from a model of the underlying QCD structure of the vacuum 
\cite{Diak86,DPPob88,mitja:rev,instant1:rev};

\item  obtained by modeling the effective gluon propagator and using
Schwinger-Dyson techniques \cite{Roberts94,Roberts:rev};

\item  derived from QCD  inspired models \cite
{Eguchi:boson,Ball87,Ball89,Ball90}.
\end{itemize}

For extensive reviews with a focus on solitons, and numerous additional
references, see \cite{mitja:rev,Alkofer:rev,ripka:book}. The four-quark
interaction leads to the chiral symmetry breaking, which is, in the
framework of these models, the key dynamical ingredient.

The various models fall into two categories: {\em local} models, where the
four-quark interaction is point-like and where ultra-violet divergences are
removed by introducing cut-offs in the quark loop, and {\em nonlocal}
models, where the four-quark interaction is smeared, such that all Feynman
diagrams in the theory are finite. Nonlocal models arise naturally in
several approaches to low-energy quark dynamics, such as the
instanton-liquid model \cite{Diakonov86} or Schwinger-Dyson resummations 
\cite{Roberts94}. For the derivations and farther applications of nonlocal
quark models see, {\em e.g.},\cite
{Ball90,Ripka97,Cahill87,Holdom89,Krewald92,Ripka93,BowlerB,%
PlantB,coim99wb,bled99,%
Arrio,ArrSal,Plant00,Diak00,Gocke,Prasz}.

Considerable effort has been exerted to describe baryons as solitons of
effective chiral quark models \cite{%
Alkofer:rev,KRS84,KR84,BirBan8485,Eiskal,BB8586,McGovern,NJL:rev}. The
solitons appear as bound states of quarks. Except for our earlier-reported
work \cite{nls,Bojancoim}, solitons have so far only been obtained from
local models, which typically use the Schwinger proper-time or the
Pauli-Villars regularization of the Dirac sea \cite{Alkofer:rev,Goeke96}.
One problem encountered with the proper-time regularization, by far the most
commonly used, is that the solitons turn out to be unstable unless the
auxiliary sigma and pion fields, introduced in the process of
semi-bosonization, are constrained to lie on the chiral circle \cite
{Goeke92,Ripka93d}. Such a constraint is external to the model itself.
Nonlocal models do not suffer from this instability: we have shown \cite
{nls,Bojancoim} that stable solitons exist in a chiral quark model with
nonlocal interactions without the extra chiral-circle constraint. In fact,
we find appreciable deviations from the chiral circle. Furthermore, in local
models, the regularization is applied only to the real (non-anomalous) part
of the Euclidean quark-loop term. The finite imaginary (anomalous) part is
left unregularized in order to properly describe anomalous processes \cite
{bbs}. One may view such a division as rather artificial, and we find it
quite appealing that, with nonlocal regulators, both the real and imaginary
parts of the action are treated on equal footing. Moreover, the {\em %
anomalies are preserved} \cite{Cahill87,Holdom89,Ripka93,Arrio,ArrSal} 
and the {\em %
charges are properly quantized} \cite{nls}. Finally, the momentum-dependent
regulator makes the theory {\em finite to all orders} in the $1/N_{c}$
expansion. This is in contrast to local models, where inclusion of
higher-order-loop effects requires extra regulators \cite
{Emil,thesis,Temp,Oertel,Oertel:thesis} and the predictive power is
diminished.

The above-mentioned virtues of nonlocal models do not come for free. Two
complications arise. First, soliton calculations require an extra
integration over an energy variable, which has to be performed numerically.
Second, the Noether currents in nonlocal models acquire extra contributions,
which are, however, needed to conserve the Noether currents and anomalies.
Nevertheless, an ambiguity remains: the transverse parts of currents are not
fixed by current conservation and their choice is effectively part of the
model building. This problem has been known for a long time in nuclear
physics, where the transverse part of the meson-exchange currents is
ambiguous \cite{BowlerB,PlantB}. It is not possible to get rid of this
ambiguity in gauging nonlocal models. An ideal solution would be to first
gauge the underlying theory ({\em e.g.} the instanton model of the QCD
vacuum), and then to derive an effective gauged quark model. This has not
been attempted so far, so that in effective nonlocal quark models we need to
deal with transverse currents which are not uniquely defined. Fixing these
currents is a part of the model building.

In this paper we demonstrate the existence of solitons and we give the
description of formal and technical aspects of constructing solitons in
nonlocal chiral quark models.\ We do not aim for an accurate phenomenology
of the nucleon and the $\Delta $ resonance. It is known from the experience
gained in local models that, in addition to the mean-field approximation
applied in our work, many other effects can and should be included: the
projection of the center-of-mass motion and angular-momentum, \cite
{GolliRosina,Lubeck,Birse:proj,NeuberGoeke,Amoreira}, the rotational $%
1/N_{c} $ corrections \cite{Wakamatsu93,Michal:ga}, inclusion of other
degrees of freedom such as vector mesons \cite{BB8586,Schueren92} or
color-dielectric fields \cite{Ban:rev:nc}. These may considerably alter and
improve the results. We stress that the methods which we describe in this
paper are applicable to all models with {\em separable nonlocal four-fermion
interactions\/}
provided that it is possible to perform an analytic continuation in the 
four-momentum close to zero, as explained in Sect. \ref{sec:en}.

The existence of stable solitons in nonlocal models has been briefly
reported in Refs. \cite{nls,Bojancoim}. In the present work we describe in
greater detail how the soliton is constructed and how the valence orbit is
defined (Sect. \ref{sec:en}). The determination of the model parameters is
described in Sect. \ref{sec:vacuum}. The physical properties of the
soliton and its energetic stability are analyzed in Sect. \ref
{sec:properties}. An important and novel feature of this paper concerns the
gauging of the nonlocal model and the construction of Noether currents. In
Sect. \ref{sec:noether}, we apply the technique of path-ordered $P$%
-exponents and we obtain general expressions for currents in the soliton
background. Interesting results follow: the properties of the soliton, which
are related to currents at zero momenta (charges, $g_{A}$, and in general $n$%
-point Green functions with vanishing momenta on external legs) do not
depend on the choice of the path in the $P$-exponent, and are thus
universal. On the other hand, observables which probe the soliton at
non-zero momenta (form-factors, magnetic moments) do depend on the gauging
prescription. However, in the cases we have explored, the contributions of
the nonlocal parts of currents are small, in particular for sufficiently
large solitons. In Sect. \ref{sec:observables} we apply two prescriptions
to compute observables: straight-line paths in the $P$-exponents and the
weak-nonlocality limit, where the energy scale of the regulator is assumed
to be parametrically much larger than the other scales in the problem.

The many-body techniques applied in our work are inherited from the
experience acquired with local models. In particular, the nucleon is derived
from the hedgehog soliton by cranking. We calculate the electric isoscalar
radius, the isovector magnetic moment an the isovector magnetic radius.
These quantities do not depend dynamically on cranking \cite{CB86}.\ They
are much simpler to evaluate numerically, involving 
single spectral sums, than the
moment of inertia, isovector electric radius, or the isoscalar magnetic
properties, which involve double spectral sums.

Appendix \ref{gauge} contains a detailed account of the gauging method.
Appendix \ref{app:noether} describes the construction of general Noether
currents. Explicit forms for the straight-line $P$-exponents are given in
App. \ref{app:straight} and the forms obtained from the weak-nonlocality limit
are given in App. \ref{app:leading}. Expressions needed for the
evaluation of observables are displayed in App. \ref{app:observables}.

\section{The model}

\label{sec:model}

This section describes the model used to calculate the solitons. We derive
the basic formulas, such as the Euler-Lagrange equations, and the
expressions for the baryon number and energy. We discuss the construction of
the valence orbit, which is non-trivial in nonlocal models \cite{nls}.

\subsection{The action}

\label{sec:action}

The model is defined by the Euclidean action 
\begin{equation}
I=\int d_{4}x\;\left[ \bar{q}\left( -i\partial _{\mu }\gamma _{\mu
}+m\right) q-\frac{G^{2}}{2}\left( \bar{\psi}\Gamma _{a}\psi \right) ^{2}%
\right] \;,  \label{actalarip}
\end{equation}
where $q\left( x\right) $ is the quark field, $m$ the current quark mass, $%
\gamma _{\mu }=\gamma ^{\mu }=\left( i\beta ,\vec{\gamma}\right) $ are the
antihermitian (Euclidean) Dirac matrices, and $\Gamma _{0}=1$, $\Gamma
_{a}=i\gamma _{5}\tau _{a}$ define the coupling in the scalar and
pseudoscalar $q\bar{q}$ channels. We work with $N_{f}=2$ flavors. The
coupling constant $G$ has the dimension of inverse energy. The delocalized
quark field, $\psi \left( x\right) $, is related to the quark field $q\left(
x\right) $ by a regulator $r$, which is diagonal in the momentum space: 
\begin{equation}
\psi \left( x\right) =\left\langle x|\psi \right\rangle =\left\langle
x\left| r\right| q\right\rangle =\frac{1}{\Omega }\sum_{k}r_{k}e^{ik\cdot
x}q_{k}\;,
\end{equation}
where $\Omega =\int d_{4}x$ is the Euclidean space-time volume.

Our calculations use either a Gaussian (as first considered in Refs. \cite
{BowlerB,PlantB}) or a monopole \cite{monopole} form for the regulator: 
\begin{equation}
r_{k}=e^{-\frac{k^{2}}{2\Lambda ^{2}}}\;\;\;\left( {\rm Gaussian}\right)
,\;\;\;\;\;r_{k}=\frac{1}{1+\frac{k^{2}}{2 \Lambda ^{2}}}\;\;\;\left( {\rm %
monopole}\right) \;.  \label{regulator}
\end{equation}

The action (\ref{actalarip}) is easily cast into the form 
\begin{eqnarray}
I &=&\frac{1}{\Omega }\sum_{k}\bar{q}_{k}\left( k_{\mu }\gamma _{\mu
}+m\right) q_{k}  \nonumber \\
&&-\frac{G^{2}}{2\Omega ^{3}}\sum_{k_{1}k_{2}k_{3}k_{4}}\delta
_{k_{1}+k_{3},k_{2}+k_{4}}\;r_{k_{1}}r_{k_{2}}r_{k_{3}}r_{k_{4}}\;\left( 
\bar{q}_{k_{1}}\Gamma _{a}q_{k_{2}}\right) \left( \bar{q}_{k_{3}}\Gamma
_{a}q_{k_{4}}\right) \,.  \label{separ}
\end{eqnarray}
The regulators appearing in the four-quark interaction have a separable form
in momentum space. The separability of the quark interaction is also present
in the instanton-liquid model \cite{mitja:rev,instant1:rev}.

Soliton calculations are more easily performed by using the {\em equivalent}
bosonized form of the action (\ref{actalarip}): 
\begin{equation}
I\left( S,P\right) =-{\rm Tr}\ln \left( -i\partial _{\mu }\gamma _{\mu
}+m+r\,\Phi \,r\right) +\frac{1}{2G^{2}}\int d_{4}x\left(
S^{2}+P_{a}^{2}\right) ,  \label{action}
\end{equation}
where 
\begin{equation}
\Phi \left( x\right) =S\left( x\right) +i\gamma _{5}P_{a}\left( x\right)
\tau _{a}  \label{chiphi}
\end{equation}
is the local chiral field which is the {\em dynamical variable} of the
system. The trace in Eq. (\ref{action}) is over color, flavor, Dirac
indices, and space-time. The chiral fields represent the following
expectation values of bilinear forms of the quark fields: 
\begin{eqnarray}
S\left( x\right) &=&-G^{2}\langle \bar{\psi}\left( x\right) \psi \left(
x\right) \rangle =-G^{2}\left\langle \bar{q}\left| r\right| x\right\rangle
\left\langle x\left| r\right| q\right\rangle ,\;\;  \nonumber \\
P_{a}\left( x\right) &=&-G^{2}\langle \bar{\psi}\left( x\right) i\gamma
_{5}\tau _{a}\psi \left( x\right) \rangle =-G^{2}\left\langle \bar{q}\left|
r\right| x\right\rangle i\gamma _{5}\tau _{a}\left\langle x\left| r\right|
q\right\rangle .
\end{eqnarray}

Since stationary solitons involve time-independent $S$ and $P$ fields, it is
more useful to express the action in terms of the Dirac Hamiltonian 
\begin{equation}
h=\frac{\vec{\alpha} \cdot \vec{\nabla}}{i}+\beta m+\beta r\, \Phi \,r,
\end{equation}
bearing in mind, however, that the regulator makes the Dirac Hamiltonian
dependent (although diagonal) on the energy variable. The action (\ref
{action}) becomes 
\begin{equation}
I=-{\rm Tr}\ln \left( \partial _{\tau }+h\right) +\frac{1}{2G^{2}}\int
d_{4}x\left( S^{2}+P_{a}^{2}\right) ,  \label{actionham}
\end{equation}
where $\tau $ is the Euclidean time variable.

\subsection{The hedgehog soliton}

\label{sec:hedgehog}

Our use of the model action (\ref{actionham}) is the same as in other
hedgehog soliton calculations: we treat the $S$ and $P$ fields classically,
thus keeping the leading-order contribution in the number of colors, $N_{c}$%
. Interestingly, in nonlocal models the next-to-leading-order corrections
are found to be surprisingly small in the vacuum sector \cite
{Plant00,basz,Scoccola,Praszalowicz,Ripka00,Ripkabled}.

We seek a stationary point of the action, or, more precisely, a minimum of
the energy for time-independent chiral fields with a hedgehog shape 
\begin{equation}
S\left( \vec{x}\right) =S\left( |\vec{x}|\right) ,\quad P_{a}\left( \vec{x}%
\right) =\widehat{x}_{a}P\left( |\vec{x}|\right) ,  \label{hedge}
\end{equation}
where $\widehat{x}_{a}=x_{a}/|\vec{x}|$. Asymptotically, far from the
soliton, the fields recover their vacuum values $S\left( x\right) =M$ and $%
P_{a}\left( x\right) =0$. We shall refer to the vacuum value $M$ of the
scalar field as the {\em constituent quark mass} (see the discussion of what
we mean by the {\em quark mass} in Sect. \ref{sec:quark}). The soliton
represents a bound state of $N_{c}=3$ quarks occupying a valence orbit with
the grand spin and parity $G^{P}=0^{+}$, where the grand spin $\vec{G}=\vec{J%
}+\vec{I}$ is the sum of the total angular momentum and isospin \cite
{KRS84,BirBan8485}, together with a Dirac sea which is polarized by the
hedgehog field.

\subsection{The energy-dependent quark orbits}

\label{sec:orbits}

The regulator acts as a differential operator, $r\equiv r\left( -\partial
_{\tau }^{2}-\vec{\nabla}^{2}\right) $. It introduces a nonlocal interaction
between the quarks and the chiral fields. For time-independent chiral fields 
$\Phi $, defined in Eq. (\ref{chiphi}), the Dirac operator is diagonal in the
energy representation, 
\begin{equation}
\left( \partial _{\tau }+h\right) \left| \omega \right\rangle =\left(
i\omega +h(\omega ^{2})\right) \left| \omega \right\rangle ,
\end{equation}
where $h\left( \omega ^{2}\right) $ is the energy-dependent Dirac
Hamiltonian: 
\begin{equation}
h(\omega ^{2})=\frac{\vec{\alpha}\cdot \vec{\nabla}}{i}+\beta m+\beta
r(\omega ^{2}-\vec{\nabla}^{2})\,\Phi \,r(\omega ^{2}-\vec{\nabla}^{2}).
\label{hamomeg}
\end{equation}
The Kahana-Ripka basis \cite{KRS84} is convenient to diagonalize the Dirac
Hamiltonian $h$, since the regulator $r$ is diagonal in this basis. Indeed,
using the notation $\left| skljGG_{3}\right\rangle $ for the Kahana-Ripka
basis states, we have $r(\omega ^{2}-\vec{\nabla}^{2})\left|
skljGG_{3}\right\rangle =r(\omega ^{2}+k^{2})\left| skljGG_{3}\right\rangle $%
. In this notation, $k$ is the radial momentum of a quark quantized in a
spherical box, $l$ is the orbital momentum of the upper component, $j$ is
the total angular momentum of the quark, and $G$ is the grand spin, obtained
by adding $j$ and the isospin $1/2$.

We diagonalize the Dirac Hamiltonian (\ref{hamomeg}) for each value of $%
\omega $, thereby obtaining the energy-dependent quark orbits: 
\begin{equation}
h\left( \omega ^{2}\right) \left| \lambda _{\omega }\right\rangle
=e_{\lambda }\left( \omega ^{2}\right) \left| \lambda _{\omega
}\right\rangle .  \label{qkorbits}
\end{equation}
Observables can then be calculated in terms of the quark orbits $\left|
\lambda _{\omega }\right\rangle $. A technical complication, compared to
earlier calculations which used local models, is the presence of an
additional integral over $\omega $ in the expressions for observables, which
has to be carried numerically.

\subsection{The energy, baryon number, and valence orbit of the soliton}

\label{sec:en}

We can use the basis $\left| \omega \right\rangle \left| \lambda _{\omega
}\right\rangle $ to express the action (\ref{actionham}) in the form: 
\begin{equation}
I=-\sum_{\omega ,\lambda _{\omega }}\ln \left( i\omega +e_{\lambda }\left(
\omega ^{2}\right) \right) +\frac{1}{2G^{2}}\int d_{4}x\,\left(
S^{2}+P_{a}^{2}\right)\;.  \label{actorbit}
\end{equation}
The sum over $\lambda _{\omega }$ includes color. In the zero-temperature
limit and for time-independent fields $S$ and $P$ the energy of the system
is equal to 
\begin{equation}
E=-\int \frac{d\omega }{2\pi }\sum_{\lambda _{\omega }}\ln \left( i\omega
+e_{\lambda }( \omega ^{2}) \right) +\frac{1}{2G^{2}}\int d_{3}x\,\,\left(
S^{2}+P_{a}^{2}\right) -E_{{\rm vac}}\,,  \label{energy}
\end{equation}
where $E_{{\rm vac}}$ is the vacuum subtraction, obtained by setting $%
S\left( x\right) =M$ and $P_{a}\left( x\right) =0$. This subtraction ensures
(together with the regulator) that the energy remains finite. We integrate
the first term by parts. The boundary term cancels out when the vacuum
energy is subtracted, and we find 
\begin{equation}
E=\int \frac{\omega d\omega }{2\pi }\sum_{\lambda _{\omega }} \frac{1-i\frac{%
de_{\lambda }}{d\omega }}{\omega -ie_{\lambda }\left( \omega ^{2}\right) }+%
\frac{1}{2G^{2}}\int d_{3}x\,\,\left( S^{2}+P_{a}^{2}\right) -E_{{\rm vac}%
}\,.  \label{solenerg}
\end{equation}

The determination of the baryon number of the soliton is not obvious because
the regulators appearing in the action (\ref{actalarip}) prevent us from
performing a canonical quantization of the quark field. In spite of this,
the baryon number turns out to be correctly quantized. A simple way to see
this consists in constructing the Noether current associated to the abelian
gauge transformation $q\left( x\right) \rightarrow e^{-i\eta \left( \tau
\right) }q\left( x\right) $ of the quark fields \cite{ripka:book}. For the
sake of calculating the baryon number we assume that $\eta \left( \tau
\right) $ depends only on time (complete Noether currents are derived in
App. \ref{app:noether}). The {\em quark loop} term of the action is
transformed to 
\begin{equation}
I_{q}\left( \eta \right) \rightarrow -{\rm Tr}\ln e^{i\eta }\left( \partial
_{\tau }+h\right) e^{-i\eta }=-{\rm Tr}\ln \left( -i\eta ^{\prime }+e^{i\eta
}he^{-i\eta }\right) ,  \label{iq}
\end{equation}
where $\eta ^{\prime }\equiv \partial \eta /\partial \tau $. In the Noether
construction of the baryon number extra terms arise due to the nonlocal
regulator $r$, which does not commute with $\eta $. The Dirac Hamiltonian is
diagonal in $\omega $ and $\eta $ is diagonal in $\tau $. To first order in $%
\eta ^{\prime }$ we can use the commutator expansion 
\begin{equation}
e^{i\eta }he^{-i\eta }=h+i\left[ \eta ,h\right] =h-\eta ^{\prime }h^{\prime
},
\end{equation}
where $h^{\prime }=$ $\partial h/\partial \omega $. Therefore 
\begin{equation}
I_{q}\left( \eta \right) =I_{q}\left( \eta =0\right) -{\rm Tr}\ln \left(
\partial _{\tau }+h-i\eta ^{\prime }\left( 1-ih^{\prime }\right) \right) ,
\end{equation}
and the baryon number is 
\begin{equation}
B=\frac{1}{\beta N_{c}}\frac{\partial I\left( \eta \right) }{\partial \eta
^{\prime }}=\frac{1}{\beta N_{c}}{\rm Tr}\frac{i}{\partial _{\tau }+h}\left(
1-ih^{\prime }\right) \;,
\end{equation}
where the factor $1/N_{c}$ is due to the fact that the quark carries baryon
number $1/N_{c}$. Using the spectral basis $\left| \omega \right\rangle
\left| \lambda _{\omega }\right\rangle $ to evaluate the trace, we obtain
the following expression for the baryon number of the system: 
\begin{equation}
B=\frac{1}{N_{c}}\int_{-\infty }^{\infty }\frac{d\omega }{2\pi }%
\sum_{\lambda _{\omega }}\frac{1-i\frac{de_{\lambda }(\omega )}{d\omega }}{%
\omega -ie_{\lambda }(\omega )}\;.  \label{barnum}
\end{equation}
The presence of the factors $1-i\frac{de_{\lambda }}{d\omega }$ is crucial
for the quantization of the baryon number. They appear in the calculation of
all observables related to Noether currents, such as the baryon density, $%
g_{A}$, and magnetic moments, as shown in App. \ref{app:observables}.
The factors $1-i\frac{de_{\lambda }}{d\omega }$ are the {\em inverse residues%
} of the poles of the quark propagator $1/\left( \partial _{\tau }+h\right) $%
. Indeed, in the vicinity of a pole at $\omega _{0}$ we find 
\begin{equation}
\frac{1}{\omega -ie_{\lambda }(\omega )}=\frac{1}{\omega -\omega
_{0}-i\left. \frac{de_{\lambda }(\omega )}{d\omega }\right| _{\omega
_{0}}(\omega -\omega _{0})}=\frac{1}{\omega -\omega _{0}}\frac{1}{\left(
1-i\left. \frac{de_{\lambda }(\omega )}{d\omega }\right| _{\omega
_{0}}\right) }.
\end{equation}
The position of these poles corresponds to the ``on-shell'' single-quark
energies. If we were able to deform the integration contour of the energy
variable $\omega $ such as to close it at infinity above or below the real
axis, the energy would become the sum of the on-shell quark energies, and
each pole would contribute a factor of $1$ to the baryon number (strictly, a
factor $1/N_{c}$, but we construct colorless solitons by placing $N_{c}$
quarks into each orbit). The baryon number is thus effectively {\em quantized%
} despite the fact that we cannot perform a canonical quantization of the
quark fields.

We now construct the valence orbit in the nonlocal model. In our
calculations the spectrum of the Dirac orbits $\left| \lambda _{\omega
}\right\rangle $ in the hedgehog field displays a similar pattern for all
values of the Euclidean energy variable $\omega $. It consists of a Dirac
sea, which is the set of negative energy orbits and a set of positive energy
orbits, which are separated from the Dirac sea orbits by a well defined
energy gap. Within this gap, there appears a bound positive energy $%
(I+J)^{P}=$ $0^{+}$ orbit, which we call the {\em valence orbit} and which
is well separated from both the positive energy and negative energy Dirac
sea orbits. We do not observe crossings between these sets. As a result, the
baryon number (\ref{barnum}) of the polarized Dirac sea is the same as the
baryon number of the vacuum, namely $0$. In order to obtain a soliton with
the baryon number equal to $1$, it is necessary to deform the integration
contour in Eq. (\ref{barnum}) as displayed in Fig.~1, such as to
enclose the valence pole. The valence pole is the solution of the equation 
\begin{equation}
\omega =ie_{v}(\omega ^{2}).  \label{valonshell}
\end{equation}
The energy $e_{v}\left( \omega ^{2}\right) $ is the eigenvalue of the Dirac
Hamiltonian $h\left( \omega ^{2}\right) $ corresponding to the valence
orbit. Equation (\ref{valonshell}) is a non-linear equation for $\omega $
and the solution is written as $ie_{{\rm val}}$.\ It can be found
numerically without difficulty or ambiguity. We refer to $e_{{\rm val}}$ as
to the valence energy, and the corresponding valence state, $|{\rm val}%
\rangle $, satisfies the equation $h\left( -e_{{\rm val}}^{2}\right) |{\rm %
val}\rangle =e_{{\rm val}}|{\rm val}\rangle $. When the integration path of $%
\omega $ is deformed so as to include the valence orbit, as shown in Fig.~1,
the soliton acquires baryon number $B=1$, and its energy is equal to 
\begin{equation}
E=N_{c}e_{{\rm val}}+\int_{-\infty }^{\infty }\frac{\omega d\omega }{2\pi }%
\sum_{\lambda _{\omega }}\frac{1-i\frac{de_{\lambda }}{d\omega }}{\omega
-ie_{\lambda }\left( \omega ^{2}\right) } 
+\frac{1}{2G^{2}}\int d_{3}x\,\,\left( S^{2}+P_{a}^{2}\right) -E_{{\rm vac}%
}\;.  \label{solenerg2}
\end{equation}
The first term is the valence orbit contribution, the second term, {\em with
the integration carried along the real axis} in the $\omega $ complex plane,
is the Dirac-sea contribution.\ The third term is the energy of the meson
fields.

\begin{figure}[t]
\vspace{0mm} \epsfxsize = 6.5 cm \centerline{\epsfbox{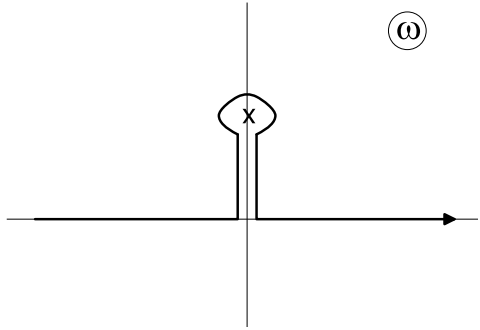}} \vspace{0mm} 
\label{contour}
\caption{The contour in the integral over the Euclidean energy variable, $%
\omega $, encircling a positive-energy valence orbit.}
\end{figure}

In addition to the valence pole, located on the imaginary
Euclidean energy axis in the vicinity of 0, there are many other poles in
the complex-energy plane. This feature is present in nonlocal models, and
also in certain local models, {\em e.g.} those with the proper-time
regularization \cite{analyt}. Fortunately, for the chosen regulators (\ref
{regulator}) the other poles lie far away from the origin on the scale $%
\Lambda $, hence they do not interfere with the valence orbit, and
the deformation of the $\omega $-integration contour
is well-defined. 


The pole defined by Eq. (\ref{valonshell}) occurs at an imaginary $%
\omega $ and therefore at a negative $\omega ^{2}$. The determination of the
valence orbit requires therefore an analytic continuation of the regulator
to negative arguments because $-e_{{\rm val}}^{2}+\vec{k}^{2}$ can become
negative for low-enough $\vec{k}^{2}$. Such analytic continuation causes no
problem for the forms (\ref{regulator}) of the regulator, but it prevents us
from using the regulator derived from the instanton model of the vacuum \cite
{Diakonov86}, because the latter has a branching point at $k^{2}=0$.
The analytic continuation is only required for small negative
values of $k^{2}=\omega ^{2}+\vec{k}^{2}$, considerably smaller than the
nonlocality scale $\Lambda ^{2}$. Indeed, $e_{{\rm val}}$ lies within $\pm M 
$, and, on the average, $k^{2}=-e_{v}^{2}+\vec{k}^{2}$ remains very small
relative to $\Lambda ^{2}.$

\subsection{The Euler-Lagrange equations for the soliton}

\label{sec:EL}

The static meson fields are determined self-consistently by solving the
Euler-Lagrange equations derived by minimizing the energy (\ref{energy})
with respect to variations of the fields $S$ and $P$:\footnote{%
The Euler-Lagrange equations in Ref. \cite{nls} had a typographical error by
which the residue factors $z_{\rm val}$
appearing in Eqs. (\ref{eulag}) were omitted.}

\begin{eqnarray}
\frac{S\left( \vec{x}\right) }{G^{2}} &=&N_{c}{\rm z}_{{\rm val}}\langle 
{\rm val}|r|\vec{x}\rangle \beta \langle \vec{x}|r|{\rm val}\rangle
+\int_{-\infty }^{\infty }\frac{d\omega }{2\pi }\sum_{\lambda _{\omega }}%
\frac{\langle \lambda _{\omega }|r|\vec{x}\rangle \beta \langle \vec{x}%
|r|\lambda _{\omega }\rangle }{i\omega +e_{\lambda }\left( \omega
^{2}\right) },  \label{eulag} \\
\frac{P_{a}\left( \vec{x}\right) }{G^{2}} &=&N_{c}{\rm z}_{{\rm val}}\langle 
{\rm val}|r|\vec{x}\rangle i\beta \gamma _{5}\tau _{a}\langle \vec{x}|r|{\rm %
val}\rangle +\int_{-\infty }^{\infty }\frac{d\omega }{2\pi }\sum_{\lambda
_{\omega }}\frac{\langle \lambda _{\omega }|r|\vec{x}\rangle i\beta \gamma
_{5}\tau _{a}\langle \vec{x}|r|\lambda _{\omega }\rangle }{i\omega
+e_{\lambda }\left( \omega ^{2}\right) },  \nonumber
\end{eqnarray}
where $\left| {\rm val}\right\rangle $ is the valence orbit, and the residue
factor is 
\begin{equation}
{\rm z}_{{\rm val}}=\left( 1-i\left. \frac{de_{{\rm val}}(\omega )}{d\omega }%
\right| _{\omega =ie_{{\rm val}}}\right) ^{-1}.
\end{equation}
Because the energies $e_{\lambda }\left( \omega ^{2}\right) $ and the wave
functions $\left| \lambda _{\omega }\right\rangle $ depend only on $\omega
^{2}$, we may symmetrize the integrands with respect to $\omega $ and
rewrite the Dirac-sea parts of Eqs. (\ref{eulag}) in the manifestly
real form
\begin{eqnarray}
&&\int_{-\infty }^{\infty }\frac{d\omega }{2\pi }\sum_{\lambda _{\omega }}%
\frac{e_{\lambda }\left( \omega ^{2}\right) \langle \lambda _{\omega }|r|%
\vec{x}\rangle \beta \langle \vec{x}|r|\lambda _{\omega }\rangle }{\omega
^{2}+e_{\lambda }^{2}\left( \omega ^{2}\right) },  \label{eulag1} \\
&&\int_{-\infty }^{\infty }\frac{d\omega }{2\pi }\sum_{\lambda _{\omega }}%
\frac{e_{\lambda }\left( \omega ^{2}\right) \langle \lambda _{\omega }|r|%
\vec{x}\rangle i\beta \gamma _{5}\tau _{a}\langle \vec{x}|r|\lambda _{\omega
}\rangle }{\omega ^{2}+e_{\lambda }^{2}\left( \omega ^{2}\right) }. 
\nonumber
\end{eqnarray}
The soliton is calculated iteratively. An initial guess is made for the
fields $S$ and $P$. In terms of these, the quark orbits (\ref{qkorbits}) are
calculated for different values of $\omega $. The fields $S$ and $P$ are
then recalculated using (\ref{eulag}), and so on until convergence is
reached.

\section{The vacuum sector}

\label{sec:vacuum}

The vacuum sector of our model, analyzed in Refs. \cite
{BowlerB,PlantB,coim99wb}, is used to constrain the model parameters, namely
the strength of the quark-quark interaction $G$, the cut-off $\Lambda $, and
the current quark mass $m$. We fit the pion decay constant, $f_{\pi }=93$%
~MeV, and the pion mass, $m_{\pi }=139$~MeV. This leaves one undetermined
parameter, which we choose to be $M$, the vacuum value of the scalar field $%
S $. Admittedly, it is somewhat abusive to claim that the model depends only
on three parameters. It is only true to the extent that the regulator $r_{k}$
is expressed in terms of one parameter, such as the cut-off appearing in (%
\ref{regulator}). The model depends in fact on the {\em shape} of the
regulator and furthermore, as we have seen in Sect. \ref{sec:en}, the
construction of the valence orbit relies on the 
possibility of the analytic continuation of
the regulator to (small) negative momenta $k^{2}$.

For smooth regulators it is found that quantities such as masses, $f_{\pi }$%
, or $m_{\pi }$, which would diverge logarithmically with the cut-off, do
not depend very much on its shape. When one of these quantities is fixed, $%
f_{\pi }$ for example, the masses of mesons and of solitons do not depend
very much on the form of the regulator. However, quantities, such as the
quark condensate, which diverge quadratically with the cut-off, are more
sensitive to the shape of $r_{k}$ and different values are obtained with
different shapes even when $f_{\pi }$ is kept fixed \cite{Ripka89}. This can
be seen from Table~\ref{table:regulators}.
In the
whole range of values $300<M<600$, the cut-off of the Gaussian regulator
remains $1.06$ times larger than the cut-off of the monopole regulator,
within 1\% accuracy. The two regulators have the same low-$k$ behavior,
therefore 
the shape dependence is roughly a 6\% effect on quantities such as $%
f_{\pi }$ or $m_{\pi }$.

One may rephrase the above statements as follows: for any given $M$ we fit $%
\Lambda $ for each regulator (Gaussian and monopole) in such a way that $%
f_{\pi }$ is fixed to 93~MeV. Then, the regulators (\ref{regulator}), each
with its own $\Lambda $, are very similar functions of the $k^{2}$ variable
up to $k^{2}\simeq 7\;{\rm fm}^{-2}$ when $M=300$~MeV, and $k^{2}\simeq 3.5\;%
{\rm fm}^{-2}$ when $M=600$~MeV. Therefore, the quantities dominated by low
values of $k^{2}$ depend only weakly on the regulator.

\begin{table}[tb]
\centering
\begin{tabular}{|l|c|r|c|c|r|c|}
\hline\hline
Regulator & $M$ & $\Lambda $ & $m$ & $\langle {\frac{1}{2}}\bar{q}%
q\rangle^{1/3}$ & $1/G$ & $\langle {\frac{\alpha_s}{\pi}}G_{\mu \nu}^a
G^{\mu \nu}_a \rangle^{1/4}$ \\ 
& [MeV] & [MeV] & [MeV] & [MeV] & [MeV] & [MeV] \\ \hline
& 300 & 760 & 7.62 & $-222$ & 182 & 327 \\ 
& 350 & 627 & 10.4 & $-200$ & 140 & 310 \\ 
Gaussian & 400 & 543 & 13.2 & $-185$ & 113 & 298 \\ 
& 450 & 484 & 15.9 & $-174$ & 94 & 287 \\ 
& 600 & 380 & 24.2 & $-151$ & 61 & 268 \\ \hline
& 300 & 718 & 3.96 & $-276$ & 204 & 347 \\ 
& 350 & 590 & 5.24 & $-252$ & 159 & 334 \\ 
monopole & 400 & 509 & 6.44 & $-235$ & 130 & 323 \\ 
& 450 & 452 & 7.56 & $-223$ & 110 & 315 \\ 
& 600 & 352 & 10.5 & $-200$ & 74 & 294 \\ \hline\hline
\end{tabular}
\caption{The vacuum properties for the two regulators considered, and for
various values of the constituent quark mass $M$. For each case the values
of $F_\pi$ and $m_\pi$ have been fixed to
their physical values.}
\label{table:regulators}
\end{table}

\subsection{The quark propagator in the vacuum}

\label{sec:quark}

In the vacuum sector, the fields acquire the values $S=M$ and $P=0$, and the
inverse quark propagator is diagonal in momentum space. In the chiral limit $%
m\rightarrow 0$, it is equal to $k_{\mu }\gamma _{\mu }+r_{k}^{2}M$. Poles
of the quark propagator occur when $k^{2}=-M^{2}r^{4}\left( k^{2}\right) $,
where\ $k^{2}=\omega ^{2}+\vec{k}^{2}$. The solution of this equation can be
written in the form $k^{2}=-M_{q}^{2}$, where $M_{q}$ satisfies the
equation: 
\begin{equation}
M_{q}^{2}=M^{2}r^{4}\left( -M_{q}^{2}\right) .  \label{onshellqkmass}
\end{equation}
This non-linear equation for $M_{q}^{2}$ has, in general, solutions which
are scattered in the complex $k^{2}$ (or $M_{q}^{2}$) plane. If $M/\Lambda $
is small enough, a solution occurs on the real axis of the $M_{q}^{2}$ plane
and such a pole represents an on-shell free quark with a mass equal to $%
M_{q} $, which we call the {\em vacuum on-shell quark mass}. This on-shell
quark mass $M_{q}$ can be considerably larger than $M$, which we call the 
{\em constituent quark mass}. As we shall see in Sect. \ref{sec:properties},
it is the on-shell quark mass $M_{q}$, and not the constituent quark mass $M$%
, which determines the stability of the soliton (see the discussion of
Fig.~2). When the energy (\ref{solenerg2}) is greater than $%
N_{c}M_{q}$, the soliton is not formed, which means that the Euler-Lagrange
equations (\ref{eulag}) do not have a localized stationary solution. The
soliton can thus be viewed as a bound state. The stability of the soliton is
discussed at the end of Sect. \ref{sec:properties}.

\subsection{The gap equation and the condensates}

\label{sec:gap}

When discussing the model in the vacuum sector, it is much simpler to use
the Lorentz-invariant form (\ref{action}) of the action. In the vacuum
sector, a translationally invariant stationary point of the action exists
with $S=M$, $P_{a}=0$, where $M$ is the solution of the equation: 
\begin{equation}
\frac{1}{G^{2}}=4N_{c}N_{f}\int \frac{d_{4}k}{\left( 2\pi \right) ^{4}}\;%
\frac{r_{k}^{4}}{k^{2}+\left( m+r_{k}^{2}M\right) ^{2}}\;.  \label{gap}
\end{equation}
Equation (\ref{gap}) is traditionally called the ``gap equation''. It is
the Euler-Lagrange equation expressed for a translationally-invariant system
(without valence quarks). For a given current quark mass $m$ (which is
determined by fitting $m_{\pi }$), the gap equation (\ref{gap}) relates $M$
to the interaction strength $G$. In this work we use it in
order to eliminate the parameter $G$ in favor of $M$.

The quark condensate $\left\langle \bar{q}q\right\rangle =\left\langle \bar{u%
}u+\bar{d}d\right\rangle $ can be obtained from the action (\ref{action}%
). In the chiral limit it is equal to 
\begin{equation}
\left\langle \bar{q}q\right\rangle =\frac{1}{\Omega }\left. \frac{\partial I%
}{\partial m}\right| _{m=0}=4N_{c}N_{f}\int \frac{d_{4}k}{\left( 2\pi
\right) ^{4}}\;\frac{r_{k}^{2}}{k^{2}+\left( r_{k}^{2}M\right) ^{2}}.
\label{qqbar}
\end{equation}

In the instanton model of the QCD vacuum the gluon condensate can be
expressed in terms of the constituent quark mass and the four-fermion
coupling constant \cite{Weiss}: 
\begin{equation}
\left\langle \frac{\alpha _{s}}{\pi }G_{\mu \nu }^{a}G_{a}^{\mu \nu
}\right\rangle =32N_{c}\int \frac{d_{4}k}{(2\pi )^{4}}\frac{r_{k}^{4}M^{2}}{%
k^{2}+\left( r_{k}^{2}M\right) ^{2}}=\frac{8M^{2}}{N_{f}G^{2}}.
\label{gg}
\end{equation}
The numerical values are listed in Table~\ref{table:regulators}. The
estimate for the gluon condensate inferred from the QCD sum rules \cite
{QCD:sum:1,QCD:sum:review} is $\left\langle \frac{\alpha _{s}}{\pi }%
G^{2}\right\rangle ^{1/4}=360\pm 20\,{\rm MeV}$, while Ref. \cite{Narison95}
gives the value $386\pm 10\,{\rm MeV}$. These estimates, when compared to
the numbers in Table~\ref{table:regulators}, favor lower values of $M$ in
our model.

\subsection{The pion mass and the pion decay constant}

\label{sec:fpi}

The inverse pion propagator $K_{P}^{-1}(x,y)$ in the vacuum can be deduced
from the action (\ref{action}): 
\begin{equation}
\left\langle xa\left| K_{P}^{-1}\right| yb\right\rangle =\frac{\delta
^{2}I\left( S,P\right) }{\delta P_{a}\left( x\right) \delta P_{b}\left(
y\right) }\;.
\end{equation}
Explicit expressions for the meson propagators, $f_{\pi }$, and $m_{\pi }$,
calculated with the present model, can be found in Refs. \cite{%
Diakonov86,BowlerB,PlantB,coim99wb,Ripka00} 
and we shall not reproduce them here. We
only specify the expressions used to determine the model parameters. The
pion propagator is diagonal in momentum and flavor space: $\left\langle
qa\left| K_{P}^{-1}\right| q^{\prime }b\right\rangle =\delta _{ab}\delta
_{qq^{\prime }}K_{P}^{-1}\left( q^{2}\right) $. To lowest order in $q^{2}$
and $m$ it acquires the form 
\begin{equation}
K_{P}^{-1}\left( q^{2}\right) =Z_{\pi }\left( q^{2}+m_{\pi }^{2}\right) ,
\end{equation}
where the inverse residue of the pion pole is 
\begin{equation}
Z_{\pi }=2N_{c}N_{f}\;f\left( M\right)
\end{equation}
and where the function $f\left( M\right) $ is 
\begin{equation}
f\left( M\right) =\int \frac{d_{4}k}{\left( 2\pi \right) ^{4}}\frac{%
r_{k}^{4}-k^{2}r_{k}^{2}\frac{dr_{k}^{2}}{dk^{2}}+k^{4}\left( \frac{%
dr_{k}^{2}}{dk^{2}}\right) ^{2}}{\left( k^{2}+r_{k}^{4}M^{2}\right) ^{2}}\;.
\label{fm}
\end{equation}
The pion decay constant, $f_{\pi }$, is then given by the expression 
\cite{Diakonov86,BowlerB,PlantB} 
\begin{equation}
f_{\pi }^{2}=Z_{\pi }M^{2}=2N_{c}N_{f}M^{2}\;f\left( M\right) ,  \label{fpi}
\end{equation}
and the pion mass equals to \cite{Diakonov86,BowlerB,PlantB} 
\begin{equation}
m_{\pi }^{2}=\frac{2m}{M\;f\left( M\right) }g^{\prime }\left( M\right)
,\;\;\;\;g^{\prime }\left( M\right) =\int \frac{d_{4}k}{\left( 2\pi \right)
^{4}}\;\frac{r_{k}^{2}}{k^{2}+r_{k}^{4}M^{2}}.  \label{mpi2}
\end{equation}
From Eq. (\ref{qqbar}) we can see that, in the chiral limit, the quark
condensate is $\left\langle \bar{q}q\right\rangle =-4N_{c}N_{f}Mg^{\prime
}\left( M\right) $. Using (\ref{fpi}), Eq. (\ref{mpi2}) can be cast
into the form 
\begin{equation}
m_{\pi }^{2}=-\frac{m\left\langle \bar{q}q\right\rangle }{f_{\pi }^{2}\;},
\end{equation}
which is the Gell-Mann--Oakes--Renner relation, requested by the constraints
of the chiral symmetry.

The model parameters $M$, $\Lambda $, $m$, and $G$ are listed in Table~\ref
{table:regulators}. We notice that when $M$ increases, the cut-off $\Lambda $
decreases. This occurs because the pion decay constant is kept fixed at the
value $f_{\pi }=93$~MeV. For large $\Lambda $, the quark condensate
increases linearly with $M$ and decreases quadratically with $\Lambda $. The
net result is a slow decreases with $M$. The coupling constant $G$ of the
attractive quark-quark interaction increases with $M$. This is why solitons
are bound only when $M$ is large enough, as discussed in Sect. \ref
{sec:properties}. At large values, $M>450\;{\rm MeV}$, the cut-off becomes
embarrassingly small as compared to the other scales in the problem. 
The error of taking the leading expressions in $m$ extends from 2~\% 
at $M=300$~MeV to 10~\% at $M=600$~MeV.
This estimate is based on the difference of the integrals 
(\ref{qqbar}), (\ref{gg}), (\ref{fm}) and (\ref{mpi2})
in the chiral limit with  $m$ from Table~\ref{table:regulators}.

\begin{figure}[tb]
\vspace{0mm} \epsfxsize = 10.5 cm \centerline{\epsfbox{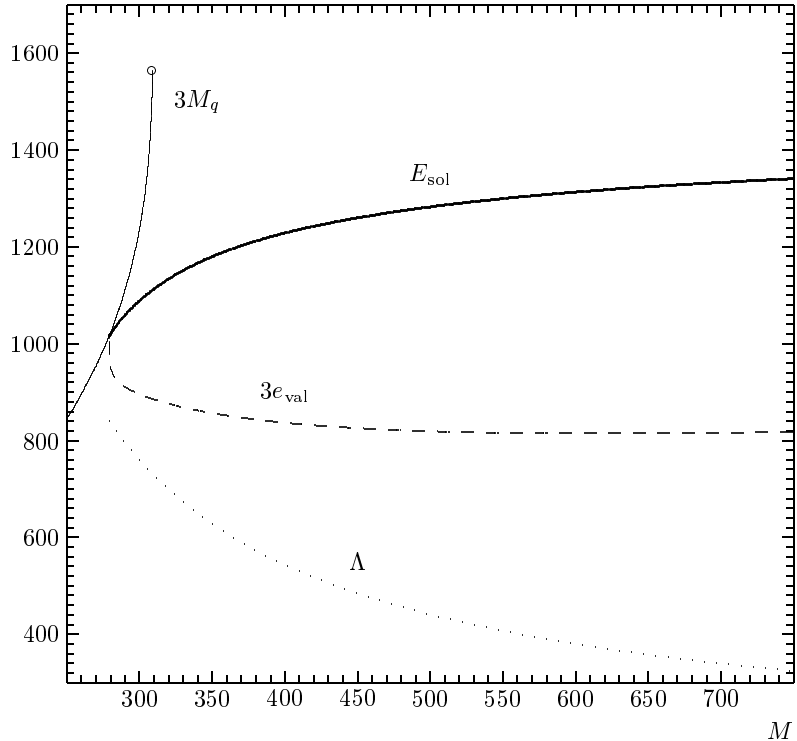}} \vspace{0mm%
} \label{fige}
\caption{The energy of the soliton (bold solid line), $N_{c}$ times the
free-space on-shell quark mass (thin solid line), and the valence
contribution to the soliton energy (dashed line), plotted as functions of
the parameter $M$. The dotted line shows the cut-off parameter, $\Lambda $,
fitted to obtain $f_\pi =93$~MeV. All quantities in MeV. The
Gaussian regulator is used.}
\end{figure}

\section{Properties of the soliton}

\label{sec:properties}

The soliton is calculated self-consistently by solving iteratively the
Euler-Lagrange equations (\ref{eulag}) for the meson fields and the Dirac
equation for the quark orbits (\ref{qkorbits}), as described in Sect. \ref
{sec:model}. The convergence is fast except for very low values of $M$.
Because of the presence of the regulator which cuts off very high momenta, 
it is
not necessary to use a very large basis as in similar calculations in
local models. We have performed the calculation with two shapes (\ref
{regulator}) of the regulator, Gaussian and the monopole (\ref{regulator}).
The soliton properties are sensitive to low values of $k^{2}$, such that,
according to the discussion in the beginning of Sect. \ref{sec:vacuum}, the
results depend very weakly on the shape of the regulator \cite{boj99} (see
Table~\ref{table:energies}) \footnote{%
This is true for regulators which, as a function of $k^2$, have
non-zero slope at the origin.
No stable solutions are found for regulators 
for which this derivative is zero.
}.

\begin{figure}[tb]
\vspace{0mm} \epsfxsize = 10.5 cm \centerline{\epsfbox{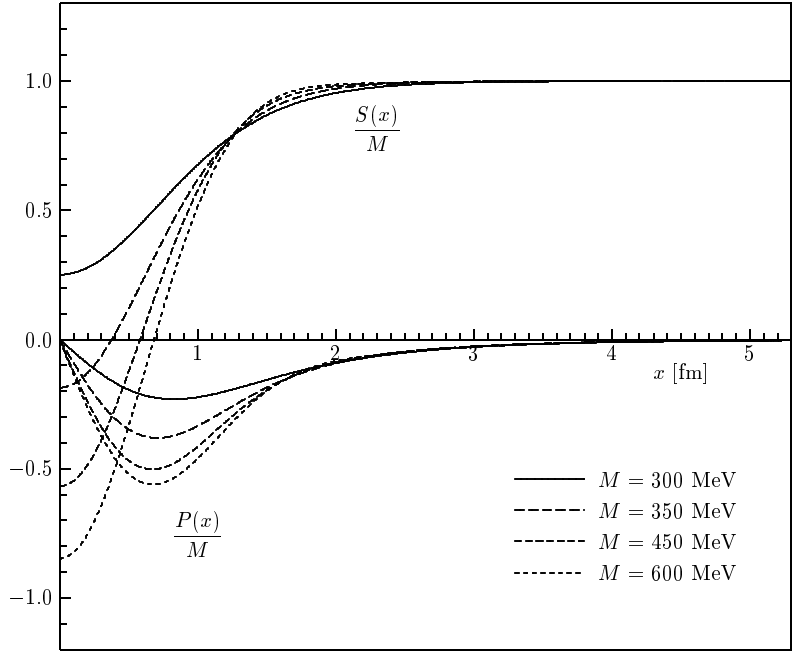}} \vspace{%
0mm} \label{figmf}
\caption{Self consistently determined fields for various values of $M$,
plotted as functions of the radial coordinate $x$. The Gaussian regulator is
used.}
\end{figure}

Figure~2 shows the energy of the soliton obtained with the Gaussian
regulator. All quantities are plotted as a function of the constituent quark
mass, $M$. The energy of the soliton is a slightly increasing function of $M$
while the energy of the valence orbit flattens out. The soliton energy
varies from about 1100 to 1250~MeV when $M$ increases from 300~MeV to
450~MeV. These seem to be reasonable values for a soliton which is to
describe the nucleon, because the energy is expected to decrease by about
200~MeV when the center-of-mass energy is subtracted in a suitable
projection method \cite{GolliRosina,Lubeck,Birse:proj,NeuberGoeke,Amoreira}.

The curve labelled $3M_{q}$ gives $N_{c}=3$ times the value of the {\em %
on-shell} quark mass $M_{q}$ in the vacuum. It is a solution of Eq. (%
\ref{onshellqkmass}). The curve terminates when $M\simeq 309$~MeV, or, more
precisely, when $\frac{M}{\Lambda }\simeq \sqrt{\frac{1}{2e}}=0.43$. (The
exact value depends on the shape of the regulator and on $m$.) Indeed, for
larger values of $\frac{M}{\Lambda }$, Eq. (\ref{onshellqkmass}) no
longer has real solutions and the ``on-shell'' quark mass wanders off into
the complex plane. 
This has been claimed to be related to quark confinement \cite
{Roberts94,Krewald92},
although the nature of this relation is far from clear.
The on shell valence orbit in the background hedgehog field, 
defined by Eq. (\ref{valonshell}), exists even when the model
parameters prevent the occurrence of an on-shell quark pole in
the vacuum background field.

A bound state of quarks occurs when the soliton energy $E$ is lower than the
energy of $N_{c}=3$ on-shell quarks, that is, when $E<N_{c}M_{q}$. Figure~2
shows that a bound state of quarks only occurs when $M$ exceeds a critical
value of 276~MeV ({\em i.e.} when the coupling constant $G>4.12/\Lambda $ ).
Beyond the critical value of $M$ where $M_q$ becomes complex, a stable
solution continues to exist\footnote{%
Since complex poles cannot be asymptotic states of the theory, the
soliton (or any other object) can never decay into such states.}.
A very similar behavior is found for the monopole regulator.

\begin{figure}[tb]
\vspace{0mm} \epsfxsize = 10.5 cm \centerline{\epsfbox{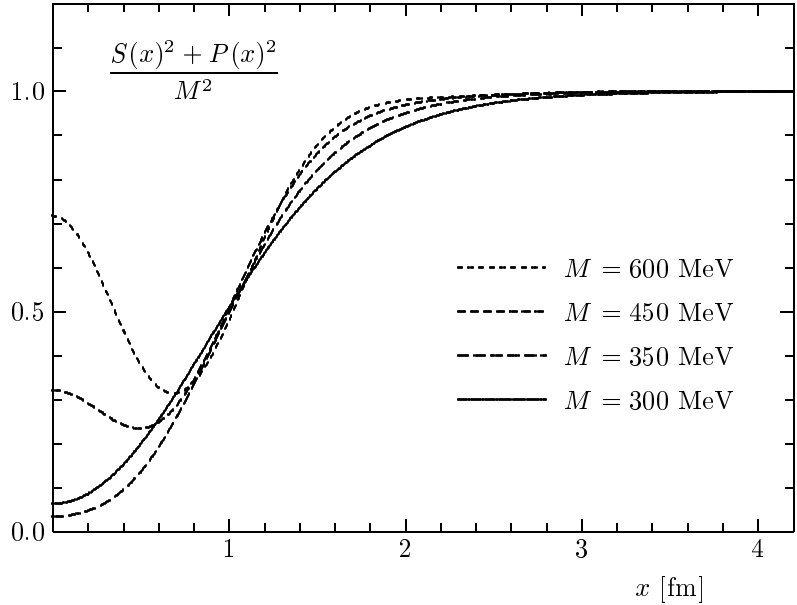}} \vspace{%
0mm} \label{figphi2}
\caption{Effective squared quark mass in the soliton plotted as a function
of the radial coordinate $x$. The Gaussian regulator is used.}
\end{figure}

Figure~3 shows the radial shapes of the hedgehog chiral fields $S\left(
x\right) $ and $P\left( x\right) $ in units of $M$. They are the solutions
of the Euler-Lagrange equations (\ref{eulag}). We can see that, in the
reasonable range $350\;{\rm MeV}<M<450\;{\rm MeV}$, the chiral field
deviates significantly from the chiral circle. Only at excessively high
values, $M>600$~MeV, does the chiral field remain close to the chiral circle.
This is a new dynamical result. In previous soliton calculations, which used
the Nambu Jona-Lasinio model with proper-time or Pauli-Villars
regularizations \cite{Alkofer:rev,Goeke96}, the soliton was found to be
unstable (the energy was unbounded from below) unless the fields were
artificially constrained to the chiral circle \cite{Goeke92,Ripka93d}. Our
model is free from this restriction. In our soliton, the self-consistent
pion field is considerably smaller, as compared to previous calculations.\
In a sense, it is midway between the Skyrmion (where the chiral field is on
the chiral circle) and the Friedberg-Lee soliton \cite{F-Lee,Wilets:rev}
(where the pion field is zero). Deviations from the chiral circle are
further illustrated in Fig.~4 which shows the values of $S^{2}\left(
x\right) +P_{a}^{2}\left( x\right) $ in units of $M^{2}$. The curve would
remain constant and equal to 1 if the fields remained on the chiral circle.
Another way to phrase the behavior displayed in Fig. 4 is to say that the
chiral symmetry is partially restored in the center of the soliton.

\begin{figure}[tb]
\vspace{0mm} \epsfxsize = 10.5 cm \centerline{\epsfbox{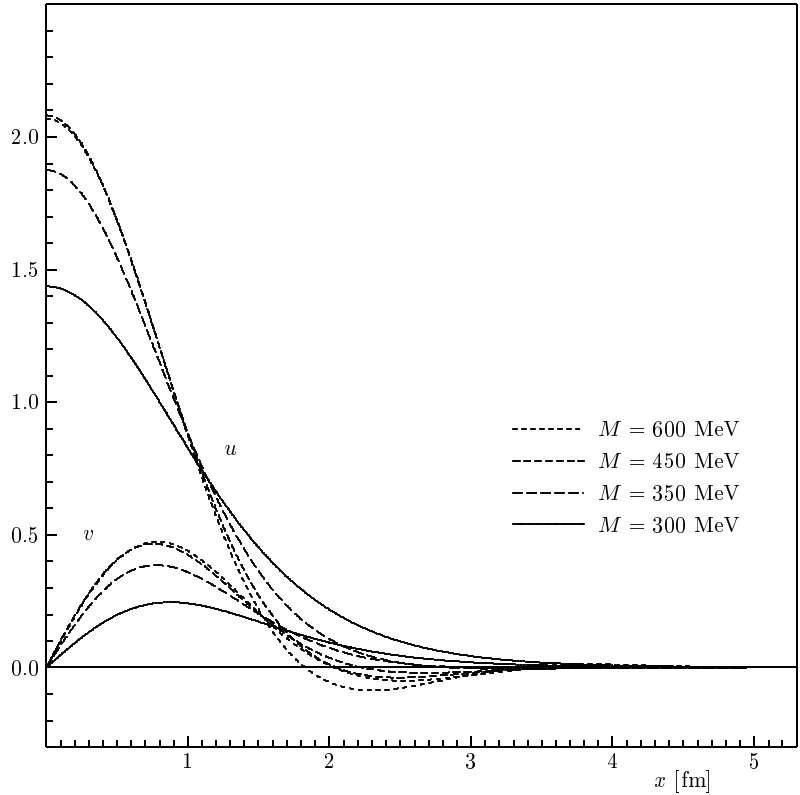}} \vspace{%
0mm} \label{figqf}
\caption{The upper ($u$) and lower ($v$) quark components for the valence
orbit for various values of $M$, plotted as functions of the radial
coordinate $x$. The Gaussian regulator is used.}
\end{figure}

Figure~5 shows the upper ($u$) and lower ($v$) quark components for the
valence orbit for various values of $M$. We note that the solution shrinks
as $M$ is increased, however, beyond $M \simeq 450$~MeV the effect saturates.

\begin{figure}[tb]
\vspace{0mm} \epsfxsize = 10.5 cm \centerline{\epsfbox{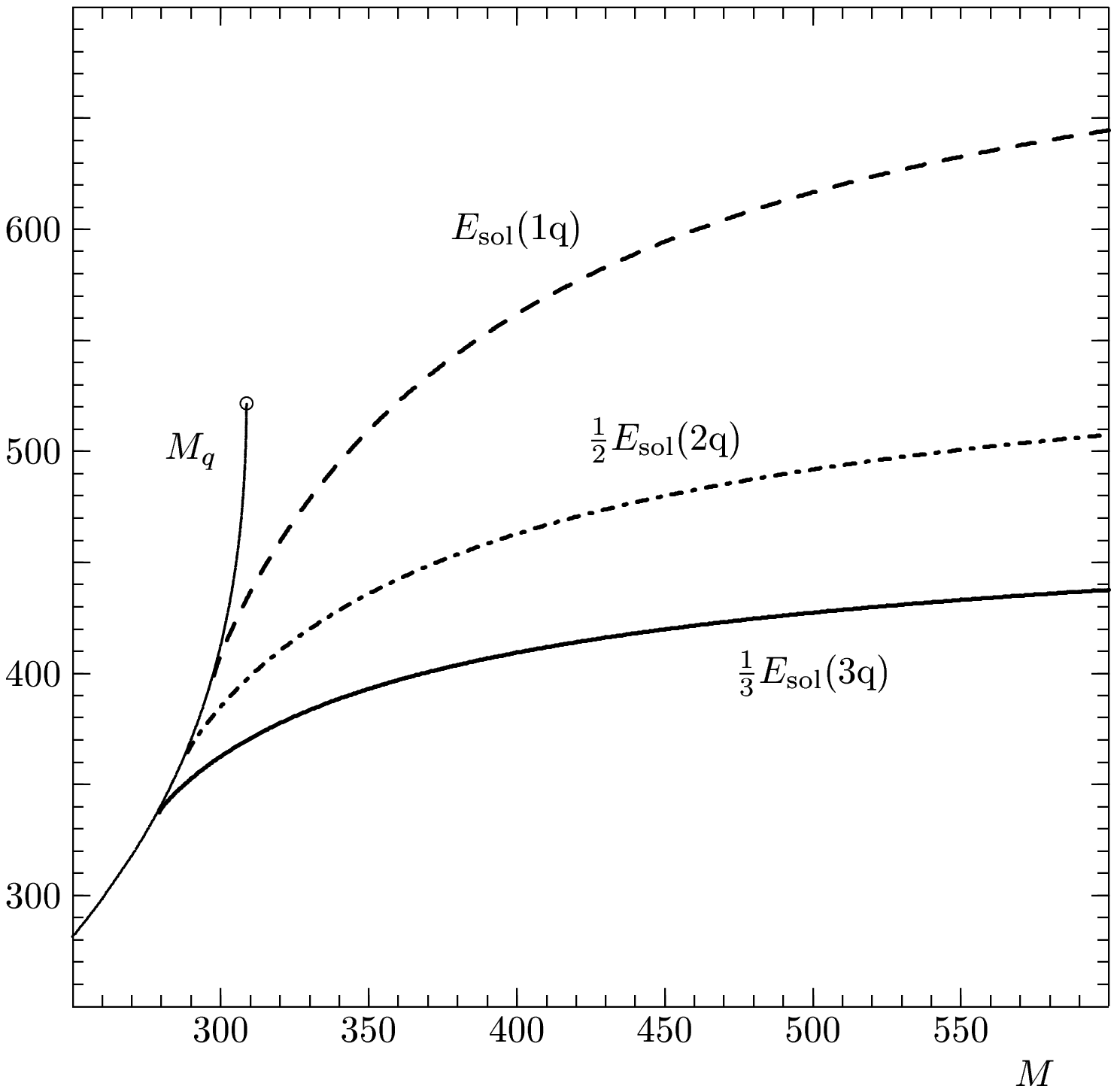}} \vspace{%
0mm} \label{figes}
\caption{The energy per quark for the soliton with three valence quarks
(bold solid line), one valence quark (dashed line), and two valence quarks
(dashed-dotted line), plotted as functions of the parameter $M$. The thin
solid line shows free-space on-shell quark mass $M_q$. All quantities in MeV.
The Gaussian regulator is used. }
\end{figure}

Figure~6 illustrates the stability of the soliton, composed of 3 quarks,
with respect to its fragmentation into solitons formed with 1 or 2 quarks.
Due to the lack of confinement, such solutions formally exist in the model.
The solid line, labelled $\frac{1}{3}E_{sol}\left( 3q\right) $, shows the
soliton energy {\em per quark}. The dashed and dot-dashed lines show the
energy of the single-quark and two-quark soliton, respectively. The curve
labelled $M_{q}$ is the on-shell mass of the quark {\em in the vacuum}. We
conclude from Fig. 6 that the 3-quark soliton is stable against the
breakup into solitons with a lower number of quarks. Similar results have
been found in the linear sigma model with valence quarks \cite{Golli97}.
Note that the Pauli principle prevents placing more than $N_{c}=3$ quarks
into the $0^{+}$ valence orbit.

\begin{table}[tb]
\centering
\begin{tabular}{|l|c|c|c|r|c|}
\hline\hline
Regulator & $M$ & $e_{{\rm val}}$ & $E_{{\rm sea}}$ & $E_{{\rm %
mes}}$ & $E_{{\rm sol}}$ \\ 
& [MeV] & [MeV] & [MeV] & [MeV] & [MeV] \\ \hline
& 300 & 298 & 2420 & $-2227$ & 1088 \\ 
& 350 & 285 & 1773 & $-1450$ & 1180 \\ 
Gaussian & 400 & 279 & 1494 & $-1102$ & 1228 \\ 
& 450 & 275 & 1339 & $-905$ & 1260 \\ 
& 600 & 272 & 1117 & $-619$ & 1313 \\ \hline
& 300 & 289 & 3008 & $-2790$ & 1084 \\ 
& 350 & 275 & 2201 & $-1850$ & 1176 \\ 
monopole & 400 & 266 & 1835 & $-1407$ & 1227 \\ 
& 450 & 260 & 1628 & $-1147$ & 1261 \\ 
& 600 & 248 & 1332 & $-753$ & 1321 \\ \hline\hline
\end{tabular}
\caption{Contributions to the soliton energy calculated with various
regulators.}
\label{table:energies}
\end{table}

\section{Noether currents}

\label{sec:noether}

The Noether currents in nonlocal models acquire extra contributions due to
the momentum-dependent regulator. As mentioned in the introduction, the
transverse parts of Noether currents are not fixed by current conservation
and their choice is not unique. Any prescription becomes an element of the
model building. The problem of ambiguous transverse currents has been known
for a long time in connection with meson-exchange currents \cite
{BowlerB,PlantB}. An elegant way of gauging the nonlocal model is to use
path-ordered $P$-exponents, and we choose this technique to construct
the Noether currents.

The details are shown in Appendices \ref{gauge}-\ref{app:noether}. In this
section we make some general remarks:

\begin{itemize}
\item  The ambiguity in the choice of Noether currents is attributed to the
freedom of choosing the path in the $P$-exponent.

\item  The Noether currents associated to symmetries are conserved.

\item  The properties of solitons involving zero-momentum probes, such as
the baryon number and $g_{A}$, do not depend on the chosen path in the
$P$-exponent.\ They are thus defined unambiguously.

\item  Soliton radii, magnetic moments, form factors, do depend on the path,
and hence they are not uniquely determined. We find, however, that the
path-dependence is weak in the weak-nonlocality limit,{\em \ i.e.} in the
case where the soliton scales are much smaller than the nonlocality scale, $%
\Lambda .$ This is the case for sufficiently large solitons.
\end{itemize}

Using a straight-line path in the $P$-exponent, we derive in App. \ref
{app:straight} the following form for the nonlocal contributions to Noether
currents: 
\begin{eqnarray}
j_{\mu a}^{{\rm NL,straight}}\left( \vec{z}\right) &=&-\int \frac{d\omega }{%
2\pi }\sum_{\lambda _{\omega }}\int d_{3}x\int d_{3}y\int_{0}^{1}d\alpha
\,\delta (\vec{z}-\vec{x}-\alpha \left( \vec{y}-\vec{x}\right) )\times 
\nonumber \\
&&\frac{\langle \lambda _{\omega }|\vec{x}\rangle \langle \vec{x}|r_{\mu
}(\omega )|\vec{y}\rangle \langle \vec{y}|\lambda _{a}\beta \Phi r(\omega
)|\lambda _{\omega }\rangle }{i\omega +e_{\lambda }\left( \omega \right) }%
+h.c.,  \label{curstr}
\end{eqnarray}
where $\lambda _{a}$ stands for $1/N_{c},\tau _{a}/2,$ or $\gamma _{5}\tau
_{a}/2$ in the case of the baryon, isospin, or axial currents, respectively.
The parameter $\alpha $ describes the straight-line integration path. The
space integrations reflect the nonlocality. The form (\ref{curstr}) looks
somewhat complicated. However, {\em moments} of currents 
($g_{A}$, radii, or magnetic moments) are very easily evaluated because,
for those cases, the $\alpha $ integration and one space integration can be
performed trivially (see App. \ref{app:observables} for examples)
leaving only one space integral, as in the local contributions.

A simpler expression for the Noether currents can be derived in the
weak-nonlocality limit \cite{ripka:book}. As shown in App. \ref
{app:leading}, we have: 
\begin{equation}
j_{\mu a}^{{\rm NL,weak}}\left( \vec{z}\right) =-\int \frac{d\omega }{2\pi }%
\sum_{\lambda _{\omega }}\frac{\langle \lambda _{\omega }|\vec{z}\rangle
\langle \vec{z}|r_{\mu }(\omega )\lambda _{a}\beta \Phi r(\omega )|\lambda
_{\omega }\rangle }{i\omega +e_{\lambda }\left( \omega \right) }+h.c.
\label{curlea}
\end{equation}

In the calculation of observables, we single out the valence and Dirac-sea
parts in Eqs. (\ref{curstr},\ref{curlea}) in the usual way, as shown in
App. \ref{app:observables}.

\section{Observables}

\label{sec:observables}

As is well known, the nucleon quantum numbers (spin, flavor) need to be
projected out of the hedgehog soliton in order to calculate observables. In
the large-$N_{c}$ limit this can be achieved by cranking \cite{CB86,ANW83}.
The observables fall into two categories, according to whether they are
dynamically-independent or dynamically-dependent on cranking. The former
ones include $g_{A}$, isoscalar-electric, and isovector magnetic properties,
while the latter ones include isovector electric and isoscalar magnetic
properties. Observables which are dynamically-independent of cranking are
simpler to evaluate, since they involve only the expectation value in the
hedgehog state. Technically, they lead to single spectral sums over the
quark orbits. Quantities which are dynamically-dependent on cranking
involve double spectral sums, and they are difficult to evaluate even in
local models \cite{Alkofer:rev,Goeke96}. The presence of nonlocality adds
extra difficulties: an integration over the energy variable, and the
appearance of nonlocal contributions. Because of these difficulties, and
since the aim in this paper is primarily to develop a consistent approach to
calculate observables in the presence of nonlocal regulators, we restrict
our calculations to the cranking-independent observables.

\subsection{The isoscalar electric radius}

\label{sec:isoel}

\begin{figure}[tb]
\vspace{0mm} \epsfxsize = 10.5 cm \centerline{\epsfbox{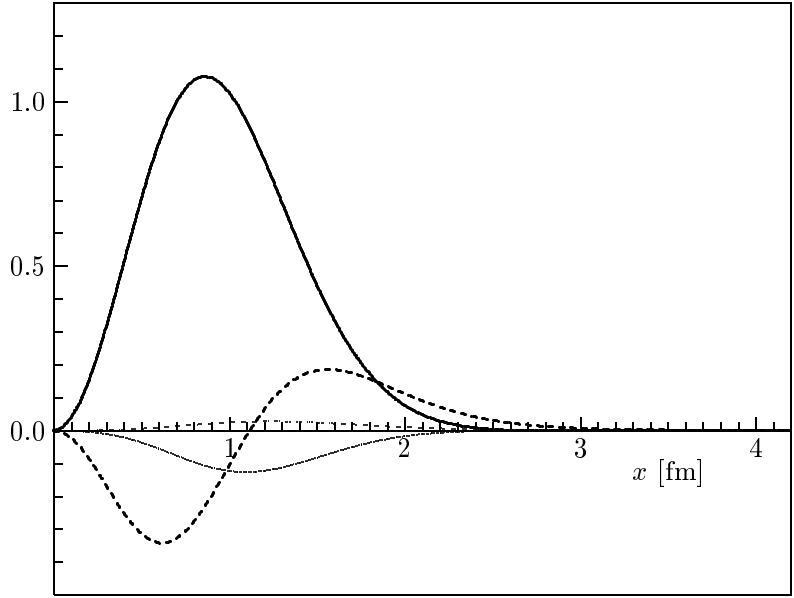}} \vspace{%
0mm} \label{figbd}
\caption{Various contributions to the baryon density (multiplied by $4 
\pi x^2$) for $M=350$~MeV and the Gaussian regulator, plotted as
functions of the radial coordinate $x$: the local (solid bold line) and
nonlocal (solid thin line) valence contributions, and the local (bold dashed
line) and nonlocal (dashed thin curve) sea contributions. The
sea contributions are multiplied by a factor of 100.}
\end{figure}

The results for the mean 
squared isoscalar electric radius, calculated with the
help of the formulas given in App. \ref{sec:ms}, are displayed in Table~%
\ref{table:electricr2} for various values of $M$. The local contribution and
different nonlocal contributions (see App. \ref{sec:ms} for their
meaning) labeled NL(A), NL(B) and NL(C), are listed separately for the
valence orbit and for the Dirac sea. As discussed in the previous section,
the radius depends on the chosen path in the $P$-exponent, and the result is
not unique. We give the results for a straight line prescription and for the
weak-nonlocality approximation.

The expressions for the nonlocal terms contain a derivative of the regulator
which produces a factor $1/(2\Lambda ^{2})$, which suppresses the nonlocal
contribution, as compared to the local one. Since $\Lambda $ decreases with
increasing $M$, the nonlocal terms become more and more important for larger 
$M$ and the difference between the two prescriptions of evaluating the
Noether currents increases. The difference is reasonably small for the
physically relevant values of $M$. The soliton is weakly bound for the
values of $M$ just above the threshold and hence very large. Its size
decreases and reaches the minimum around $M=450$~MeV. Above this value, the
radius starts increasing again. This is due to the fact that $\Lambda 
$ is very small and the size becomes proportional to the inverse $%
\Lambda $.

The nonlocal contribution from the valence orbit considerably reduces the
isoscalar electric radius, up to 20\% for low values of $M$. This is
consistent with the fact that the inverse residue, ${\rm z}_{{\rm val}}$,
and hence the integrated local density, is greater than 1. The corresponding
nonlocal density is negative for all $|\vec{x}|$ (see Fig.~7). We can also see
that the weak-nonlocality limit works better for lower values of $M$. This
is natural, because the weak-nonlocality limit may be viewed as the
leading-order term in an expansion in powers of $1/(R\Lambda )$, where $R $
describes the soliton size. This is why the $B$ terms, carrying more
derivatives of $r$, are much smaller than the $A$ and $C$ terms. By the same
argument, in the weak-nonlocality limit $A=C$. We note finally that our
numbers for $\langle r^2 \rangle_{\rm baryon}$ 
are larger than the experimental value of 0.62 fm$^{2}$.

Different contributions to the baryon (isoscalar electric) density are
displayed in Fig.~7. As a numerical check we have verified that the sum of
the valence contributions integrates to 1, while the sum of the sea
contributions integrates to 0.

\begin{table}[tb]
\centering
\begin{tabular}{|l|l|r|r|r|r|r|}
\hline\hline
\multicolumn{2}{|c|}{} & \multicolumn{5}{c|}{$M$~[MeV]} \\ \hline
\multicolumn{2}{|c|}{$\langle r^2 \rangle_{{\rm baryon}}$} & 300\  & 350\  & 
400\  & 450\  & 600 \  \\ \hline\hline
val & L & 2.209 & 1.270 & 1.057 & 0.991 & 1.062 \\ 
& NL (A) & $-0.446$ & $-0.228$ & $-0.188$ & $-0.192$ & $-0.328$ \\ 
& NL (B) & $-0.039$ & $-0.039$ & $-0.040$ & $-0.041$ & $-0.061$ \\ 
& NL (C) & $-0.435$ & $-0.194$ & $-0.119$ & $-0.074$ & 0.021 \\ 
& straight line & 1.761 & 1.052 & 0.897 & 0.852 & 0.898 \\ 
& weak NL & 1.773 & 1.075 & 0.937 & 0.917 & 1.083 \\ \hline
sea & L & 0.0050 & 0.0070 & 0.0080 & 0.0080 & 0.0075 \\ 
& NL (A) & 0.0005 & 0.0012 & 0.0020 & 0.0022 & 0.0030 \\ 
& NL (B) & 0.0001 & 0.0001 & 0.0002 & 0.0002 & 0.0003 \\ 
& NL (C) & 0.0003 & 0.0010 & 0.0015 & 0.0018 & 0.0026 \\ \hline
total & straight line & 1.766 & 1.060 & 0.907 & 0.862 & 0.909 \\ 
& weak NL & 1.778 & 1.083 & 0.947 & 0.927 & 1.094 \\ \hline\hline
\end{tabular}
\caption{Various contributions to the isoscalar electric mean square radius.
L denotes the local contribution and NL(A,B,C) different nonlocal
contributions defined in App. E.1. The total result is displayed for the
straight path and in the weak-nonlocality approximation. The Gaussian regulator
is used.}
\label{table:electricr2}
\end{table}

\subsection{$g_{A}$}

\label{sec:ga}

The results for $g_{A}$, evaluated with the expressions given in App.
\ref{app:ga}, are displayed in Table~\ref{table:ga}. The sea contribution
remains small for all values of $M$. The nonlocal terms increase with
increasing $M$, as expected, yielding, together with the local piece, almost
a constant value of $g_{A}$ over a wide range of $M$. The values, ranging
between 1.1 and 1.15, are
somewhat smaller than the experimental value of 1.26.

\begin{table}[tb]
\centering
\begin{tabular}{|l|l|l|l|l|l|}
\hline\hline
& \multicolumn{5}{c|}{$M$~[MeV]} \\ \hline
$g_A$ & \ 300 & \ 350 & \ 400 & \ 450 & \ 600 \\ \hline
val, L & 1.047 & 0.922 & 0.861 & 0.819 & 0.737 \\ 
val, NL & 0.022 & 0.069 & 0.119 & 0.170 & 0.309 \\ \hline
sea, L & 0.050 & 0.067 & 0.064 & 0.058 & 0.037 \\ 
sea, NL & 0.032 & 0.053 & 0.062 & 0.065 & 0.062 \\ \hline
total & 1.151 & 1.112 & 1.106 & 1.112 & 1.146 \\ \hline\hline
\end{tabular}
\caption{Different contributions to $g_A$ calculated from the current. L and
NL denote the local and nonlocal contributions, respectively. The Gaussian
regulator is used.}
\label{table:ga}
\end{table}

\subsection{Isovector magnetic moments and radii}

\label{sec:magnetic}

The results for the isovector magnetic moment, evaluated with the
expressions given in App. \ref{sec:mu}, are shown in Table~\ref{table:mu}%
. As in the case of $g_{A}$, the nonlocal terms increase with increasing $M$%
, and the total value is almost constant over a wide range of $M$. The
values are lower than the experimental value 4.71.

\begin{table}[tb]
\centering
\begin{tabular}{|l|l|r|r|r|r|r|}
\hline\hline
\multicolumn{2}{|c|}{} & \multicolumn{5}{c|}{$M$~[MeV]} \\ \hline
\multicolumn{2}{|c|}{$\mu_{I=1}$} & 300\  & 350\  & 400\  & 450\  & 600 \ 
\\ \hline\hline
val & L & 2.910 & 2.519 & 2.339 & 2.245 & 2.174 \\ 
& NL & 0.097 & 0.212 & 0.319 & 0.420 & 0.673 \\ \hline
sea & L & 0.293 & 0.379 & 0.386 & 0.372 & 0.305 \\ 
& NL & 0.122 & 0.198 & 0.238 & 0.262 & 0.289 \\ \hline
total &  & 3.422 & 3.307 & 3.282 & 3.299 & 3.442 \\ \hline\hline
\end{tabular}
\caption{Contributions to the isovector magnetic moment, in units of the
nuclear magneton. L and NL denote the local and nonlocal contributions,
respectively. The Gaussian regulator is used.}
\label{table:mu}
\end{table}

In Table~\ref{table:mur2} we list different contributions to the squared
isovector magnetic radius, evaluated with the expressions given in App.
\ref{sec:mu}. This quantity depends on the prescription for the Noether
current, but the difference between the straight-line path method and
the weak-nonlocality approximation is even smaller than in the case of the
baryon radius (see Table~\ref{table:electricr2}). The sea contribution is
substantially larger than in the isoscalar electric case. The numbers are
much larger than the experimental value of 0.77 fm$^{2}$.

\begin{table}[tbh]
\centering
\begin{tabular}{|l|l|r|r|r|r|r|}
\hline\hline
\multicolumn{2}{|c|}{} & \multicolumn{5}{c|}{$M$~[MeV]} \\ \hline
\multicolumn{2}{|c|}{$\langle r^2 \rangle_{M,I=1}$} & 300\  & 350\  & 400\ 
& 450\  & 600 \  \\ \hline
val & L & 2.498 & 1.288 & 1.043 & 1.001 & 1.285 \\ 
& NL & 0.073 & 0.112 & 0.153 & 0.196 & 0.334 \\ \hline
sea & L & 0.378 & 0.424 & 0.421 & 0.405 & 0.343 \\ 
& NL & 0.137 & 0.187 & 0.220 & 0.243 & 0.276 \\ \hline
total & straight line & 3.087 & 2.011 & 1.838 & 1.846 & 2.238 \\ 
& weak L & 3.087 & 2.011 & 1.837 & 1.844 & 2.226 \\ \hline\hline
\end{tabular}
\caption{Various contributions to the isovector magnetic mean square radius.
L and NL denote the local and nonlocal contributions, respectively. The
total result is displayed for the straight path and in the weak-nonlocality
approximation. The Gaussian regulator is used.}
\label{table:mur2}
\end{table}

To summarize the results for the observables, we note that the soliton, at
the present mean-field treatment, is too large. This leaves room for the
inclusion of other effects. One important effect comes from the center-of
mass corrections which reduce both the mass and the mean square radii. The
fields which describe the soliton break translational symmetry. The center
of mass of the system is not at rest and it makes a spurious contribution
both to the energy and to the mean square radii (more generally, to the form
factors). This spurious contribution should be subtracted from the
calculated values. The subtraction is a next-to-leading-order correction in $%
N_{c}$. A rough estimate can be obtained from an oscillator model. If $N_{c}$
particles of mass $m$ move in the $1s$ state of a harmonic oscillator of
frequency $\hbar \omega $, the center of mass of the system is also in a $1s$
state and it contributes $\frac{3}{4}\hbar \omega =\left\langle
P^{2}\right\rangle /2N_{c}m$ to the energy. Thus, we can correct the soliton
energies by subtracting $\left\langle P^{2}\right\rangle /2E_{sol}$ from the
calculated energy. Furthermore, in the oscillator model, the center of mass
contributes a fraction $\frac{1}{N_{c}}$ of the mean square radius, such
that one corrects the mean square radius by multiplying the calculated value
by a factor equal to $\left( 1-\frac{1}{N_{c}}\right) $. These corrections
to the soliton mass and the isoscalar electric radius may bring the
calculated values close to the experimental ones \cite{Bojancoim}. To some
extent, the too large soliton may also reflect the lack of confinement of
the model, or some other omitted dynamical factor, {\em e.g.} vector
interactions \cite{BB8586,Schueren92}. Although the chiral field is
sufficiently strong to bind the quarks, additional forces may reduce the
soliton size.

The problem with too low values of $g_{A}$ and of the isovector magnetic
moment may be resolved by rotational $1/N_{c}$ corrections, as found in
soliton calculations which use proper-time or Pauli-Villars regularizations 
\cite{Wakamatsu93,Michal:ga}. The rotational corrections produce a factor
lying between 1 and $(N_{c}+2)/N_{c}$, which brings the calculated values
closer to the experimental data. This problem requires a further study.

\subsection{Calculations with chiral fields constrained to the chiral circle}

In chiral quark models the soliton properties have so far 
only been calculated with the sigma and pion fields constrained
to lie on the chiral circle \cite{Alkofer:rev,Goeke96}.
Although no clear physical ground for such a constraint can be seen 
in the derivations of effective quark models,
it is nonetheless interesting to study its effect in our model.
\begin{figure}[tb]
\vspace{0mm} \epsfxsize = 10.5 cm \centerline{\epsfbox{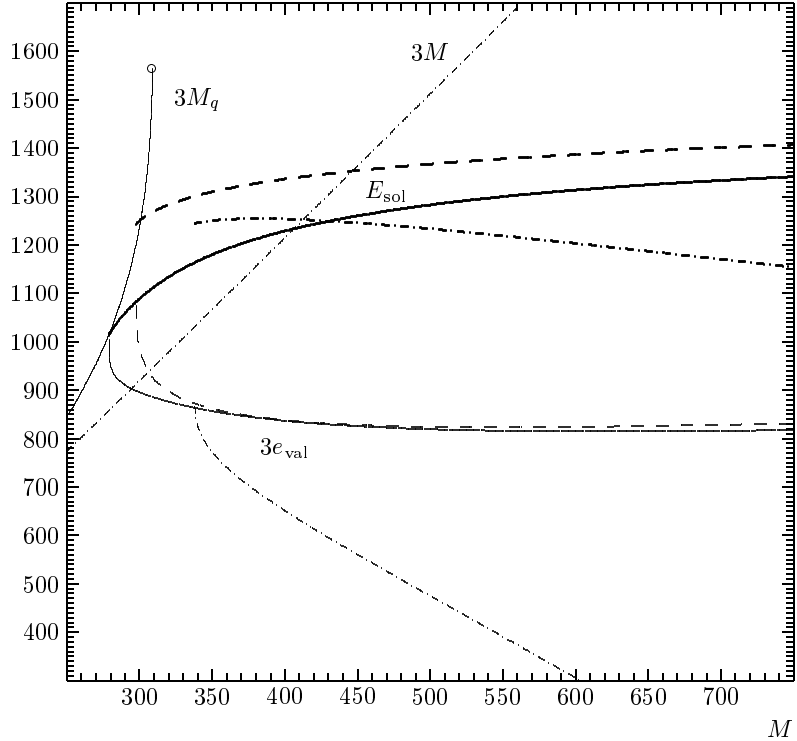}} 
\vspace{0mm} 
\label{figecc}
\caption{The energy of the soliton [in MeV] (bold lines), the three
free-space quark mass (solid line) and the valence contribution
to the soliton energy (thin lines) plotted as functions of $M$. 
Solid lines correspond to the unconstrained calculation (same as
in Fig.~2),
dashed lines to the calculation with the chiral-circle constraint, 
dashed-dotted to the calculation using proper time regularization,
also with the chiral-circle constraint.
In the latter case the energy of three free quarks is marked by $3M$.
The parameters of all models are fitted to obtain $f_\pi =93$~MeV.}
\end{figure}
Fig.~8 shows the soliton energy of the constrained calculation
as a function of $M$. 
The constrained energy is 150~MeV higher than the energy of 
our solution at smaller values of $M$ and stays some 70~MeV
higher even in the region where the 
unconstrained 
chiral fields 
come closer to the chiral circle. 
From Fig.~3 it is clear that 
in the unconstrained calculation it is energetically
favorable to increase more the magnitude of the sigma
field than the pion field.
Contrary to the unconstrained calculation where the soliton
energy at the threshold continues smoothly from the curve
representing the energy of three free quarks, the energy 
of the constrained soliton
starts abruptly at $M=298$~MeV
and in a small region of $M$ just above the threshold stays 
{\em above\/} the $3M_q$ curve.
Such an energetic instability, absent in the unconstrained nonlocal 
model, is in fact common to several chiral models \cite{%
BB86,BC86}; 
in particular, in 
the local model with the proper time regularization one finds a 
self-consistent solution for $338\;{\rm MeV} < M < 412\;{\rm MeV}$ 
even though  the soliton is energetically not stable in this region 
(see the dashed-dotted lines in Fig.~8).

The observables calculated in the constrained calculations are
displayed in Table~7.
The differences to the unconstrained results presented in
Tables~3--6 can be explained by two effects: due to a deeper
effective potential in (\ref{qkorbits}) the valence orbits
shrinks thus yielding a smaller valence contribution to
the magnetic moment and the two radii.
On the other hand, due to the stronger chiral fields, the sea
contribution to all quantities becomes more important and
partially cancels the decreased valence contribution.

\begin{table}[tb]
\centering
\begin{tabular}{|l|l|r|r|r|r|r|}
\hline\hline
\multicolumn{2}{|c|}{} & \multicolumn{5}{c|}{$M$~[MeV]} \\ \hline
\multicolumn{2}{|c|}{}      & 300\  & 350\  & 400\  & 450\  & 600 \ 
\\ \hline\hline
energy     & $e_{\mathrm{val}}$  &  327 &  287 &  279 &  276 &  275 \\
\mbox{}[MeV] & total             & 1249 & 1309 & 1336 & 1354 & 1387 \\
\hline
$\langle r^2 \rangle_{{\rm baryon}}$ 
          & valence         & 1.190 & 0.711 & 0.667 & 0.677 & 0.798 \\ 
\mbox{}[fm$^2$]  & total    & 1.201 & 0.725 & 0.684 & 0.694 & 0.813 \\ 
\hline
 $g_A$ & valence            & 1.150 & 0.978 & 0.982 & 1.003 & 1.087 \\ 
       & total              & 1.205 & 1.079 & 1.084 & 1.100 & 1.164 \\ 
\hline
$\mu_{I=1}$ & valence       & 2.444 & 2.232 & 2.259 & 2.334 & 2.647 \\ 
            & total         & 3.223 & 3.065 & 3.088 & 3.146 & 3.388 \\ 
\hline
$\langle r^2 \rangle_{M,I=1}$ & valence 
                            & 1.692 & 0.873 & 0.844 & 0.930 & 1.465 \\ 
\mbox{}[fm$^2$] & total     & 2.437 & 1.621 & 1.599 & 1.681 & 2.173 \\ 
\hline\hline
\end{tabular}
\caption{Various observables (see Tables~2--6) obtained in the 
calculation with the chiral-circle constraint.
``Valence'' stands for the sum of the local and nonlocal contributions,
the radii are calculated with the straight line prescription.}
\label{table:constrained}
\end{table}

\section{Conclusions}

\label{sec:conclude}

In this paper we have shown that a soliton, {\em i.e.} a bound state of $%
N_{c}=3$ quarks, is formed in a chiral quark model with nonlocal regulators.
We have demonstrated its energetic stability and investigated its basic
properties. Moreover, we have developed a scheme for quantizing the baryon
number of the soliton, as well as for calculating observables. The
construction of Noether currents has been accomplished through the use of
straight line path-ordered $P$-exponents. This construction is general and
applicable to any model with nonlocal separable four-fermion interactions.

We have shown that the nonlocal regularization, which is somewhat more and
yet not prohibitively complicated, has several attractive features compared
to the chiral quark models which use local regularizations, such as the
Pauli-Villars or the Schwinger proper-time method. We have found that the
pion field is considerably reduced compared to the local models which
require the chiral field to remain on the chiral circle. The soliton is
found to have properties which make it suitable for the description of the
nucleon and for application of further corrections, such as projection \cite
{GolliRosina,Lubeck,Birse:proj,NeuberGoeke,Amoreira}, or the inclusion of $%
1/N_{c}$ corrections \cite{Wakamatsu93,Michal:ga}.

The authors wish to thank Enrique Ruiz Arriola, Michael Birse, Klaus Goeke,
Maxim Polyakov, and Nikos Stefanis for many useful discussions and comments.

\appendix

\section{The gauged nonlocal model}

\label{gauge}

Consider the gauge transformation of the quark field, 
\begin{equation}
q(x)\rightarrow e^{-i\lambda ^{a}\phi _{a}(x)}q(x),  \label{gau}
\end{equation}
where $\lambda ^{a}=1/N_{c}$ for the baryon current, $\lambda ^{a}=\tau
^{a}/2$ for the isospin current, and $\lambda ^{a}=\gamma _{5}\tau ^{a}/2$
for the axial current. We gauge the nonlocal model by using the
path-ordered-exponent method described in Refs. \cite{BowlerB,PlantB,Bos}.
The method is based on the Wilson line integrals ($P$-exponents) defined as 
\begin{equation}
W_{A}(x,y)={\cal P}e^{i\int_{x}^{y}\lambda ^{a}A_{a}^{\mu }(s)ds_{\mu }},
\label{wpath}
\end{equation}
where ${\cal P}$ denotes a path ordering operator needed for non-abelian
gauge transformations, $\lambda ^{a}A_{a}^{\mu }$ is the gauge field (in
general non-abelian), and $s$ parameterizes an arbitrary path from $x$ to $y$%
. The operator $W_{A}(x,y)$ is a functional of the gauge fields $A_{a}^{\mu
} $ with the following key property: when the gauge field undergoes the
gauge transformation 
\begin{equation}
D_{\mu }\rightarrow e^{-i\lambda ^{a}\phi _{a}}D_{\mu }e^{i\lambda ^{a}\phi
_{a}},  \label{atransf}
\end{equation}
where $D_{\mu }=\partial _{\mu }+i\lambda ^{a}A_{a\mu }$, the operator $%
W_{A}(x,y)$ transforms as 
\begin{equation}
W_{A}(x,y)\rightarrow W_{A+\partial \phi }(x,y)=e^{-i\lambda ^{a}\phi
_{a}\left( x\right) }W_{A}(x,y)e^{i\lambda ^{a}\phi _{a}\left( y\right) }.
\label{watransf}
\end{equation}
Consider the {\em quark-loop} term of the Euclidean action (\ref{action}), 
\begin{equation}
I_{q}=-{\rm Tr}\ln \beta \left( -i\partial _{\mu }\gamma _{\mu }+m+r\,\Phi
\,r\right) .  \label{betact}
\end{equation}
The gauged action is constructed by making the substitutions 
\begin{equation}
\partial _{\mu }\rightarrow D_{\mu },\quad \quad \left\langle x\left|
r\right| y\right\rangle \rightarrow \left\langle x\left| r_{A}\right|
y\right\rangle =W_{A}\left( x,y\right) \left\langle x\left| r\right|
y\right\rangle ,  \label{diracrepl}
\end{equation}
and 
\begin{equation}
\beta \Phi \rightarrow e^{-i\lambda ^{a}\phi _{a}(x)}\beta \Phi e^{i\lambda
^{a}\phi _{a}(x)}.  \label{sptransf}
\end{equation}
In explicit form the gauged action reads 
\begin{equation}
I_{q}\left( A\right) =-{\rm Tr}\ln \left( -\beta i\gamma _{\mu }\partial
_{\mu }+\beta \gamma _{\mu }\lambda ^{a}A_{a\mu }+\beta m+r_{A}\beta \Phi
r_{A}\right) .  \label{iofa}
\end{equation}
Note that, in general, $r_{A}$ and $\beta $ do not commute. In the gauge
transformation the action (\ref{iofa}) transforms to 
\begin{eqnarray}
I_{q}\left( A\right) &\rightarrow &I_{q}\left( A+\partial \phi \right) = 
\nonumber \\
&&-{\rm Tr}\ln \left[ -i\beta \gamma _{\mu }\partial _{\mu }+\beta \gamma
_{\mu }\left( \lambda ^{a}A_{a\mu }+\lambda ^{a}\phi _{a\mu }+[\lambda
^{b}\phi _{b},\lambda ^{a}A_{a}^{\mu }]\right) \right.  \nonumber \\
&&\left. +\beta m+r_{A+\partial \phi }e^{-i\lambda ^{a}\phi _{a}}\beta \Phi
e^{i\lambda ^{a}\phi _{a}}r_{A+\partial \phi }\right] .
\end{eqnarray}
Using the property $r_{A+\partial \phi }=e^{-i\lambda ^{a}\phi
_{a}}r_{A}e^{i\lambda ^{a}\phi _{a}}$ we find that 
\begin{eqnarray}
I_{q}\left( A+\partial \phi \right) &=&-{\rm Tr}\ln e^{-i\lambda ^{a}\phi
_{a}}\beta \left[ -i\gamma _{\mu }\partial _{\mu }+\gamma _{\mu }\lambda
^{a}A_{a\mu }+e^{i\lambda ^{a}\phi _{a}}me^{-i\lambda ^{a}\phi _{a}}+\right.
\nonumber  \label{almostinv} \\
&&\left. \beta r_{A}\beta \Phi r_{A}\right] e^{i\lambda ^{a}\phi _{a}}.
\end{eqnarray}
Equation (\ref{almostinv}) shows that, provided $e^{i\lambda ^{a}\phi
_{a}}\beta me^{-i\lambda ^{a}\phi _{a}}=\beta m$ (the case of baryon or
isovector symmetries), or in the chiral limit $m=0$, the action is invariant
in the gauge transformation, 
\begin{equation}
I\left( A+\partial \phi \right) =I\left( A\right) ,
\end{equation}
such that the corresponding Noether current, evaluated from the expression
\begin{equation}
j_{\mu }^{a}\left( x\right) =\left. \frac{\delta I\left( A\right) }{\delta
A_{\mu }(x) }\right| _{A=0},  \label{defcur}
\end{equation}
is conserved: 
\begin{equation}
\partial _{\mu }j_{\mu }^{a}\left( x\right) =0.
\end{equation}

\section{Explicit construction of Noether currents}

\label{app:noether}

The action, expanded to first order in the $A$ field, has the form 
\begin{eqnarray}
I^{\left( A\right) } &=&-{\rm Tr}\frac{1}{-i\partial _{\mu }\beta \gamma
_{\mu }+\beta m+r\beta \Phi r}\left( \beta \gamma _{\mu }\lambda ^{a}A_{a\mu
}+r_{A}^{1}\beta \Phi r+r\beta \Phi r_{A}^{1}\right)  \nonumber \\
&=&-\sum_{\omega ,\lambda _{\omega }}\frac{\left\langle \omega ,\lambda
_{\omega }\left| \beta \gamma _{\mu }\lambda ^{a}A_{a\mu }+r_{A}^{1}\beta
\Phi r+r\beta \Phi r_{A}^{1}\right| \omega ,\lambda _{\omega }\right\rangle 
}{i\omega +e_{\lambda }\left( \omega \right) },  \label{ia}
\end{eqnarray}
where 
\begin{equation}
\left\langle x\left| r_{A}\right| y\right\rangle =\left\langle x\left|
r\right| y\right\rangle \left( 1+i\int_{x}^{y}ds_{\mu }\lambda ^{a}A_{a\mu
}\left( s\right) +...\right) \equiv \left\langle x\left| r\right|
y\right\rangle +\left\langle x\left| r_{A}^{1}\right| y\right\rangle +...
\label{ra}
\end{equation}
We can now write, with help of a representation of the $\delta $ function, 
\begin{equation}
\left\langle x\left| r_{A}^{1}\right| y\right\rangle =i\int_{x}^{y}ds_{\mu
}\lambda ^{a}A_{a\mu }\left( s\right) =i\int \frac{d_{4}q}{(2\pi )^{4}}%
\int_{x}^{y}ds_{\mu }e^{i(z-s)\cdot q}\lambda ^{a}A_{a\mu }\left( z\right) .
\label{ratrick}
\end{equation}
The Noether current is obtained from Eq. (\ref{defcur}), which,
according to (\ref{ia},\ref{ratrick}) gives a general expression, valid for 
{\em any path} in the $P$-exponent: 
\begin{eqnarray}
&&j_{\mu a}\left( z\right) \equiv j_{\mu a}^{L}\left( z\right) +j_{\mu
a}^{NL}\left( z\right)  \label{jmua} \\
&=&-\sum_{\omega ,\lambda _{\omega }}\frac{\langle \omega ,\lambda _{\omega
}\left| z\rangle \beta \gamma _{\mu }\lambda _{a}\langle z\right| \omega
,\lambda _{\omega }\rangle }{i\omega +e_{\lambda }\left( \omega \right) }%
-\sum_{\omega ,\lambda _{\omega }}i\int \frac{d_{4}q}{(2\pi )^{4}}\int
d_{4}x\int d_{4}y\int_{x}^{y}ds_{\mu }e^{i(z-s)\cdot q}\times  \nonumber \\
&&\frac{\langle \omega ,\lambda _{\omega }|x\rangle \langle x|r|y\rangle
\langle y|\lambda _{a}\beta \Phi r|\omega ,\lambda _{\omega }\rangle
+\langle \omega ,\lambda _{\omega }|r\beta \Phi \lambda _{a}|x\rangle
\langle x|r|y\rangle \langle y|\omega ,\lambda _{\omega }\rangle }{i\omega
+e_{\lambda }\left( \omega \right) },  \nonumber
\end{eqnarray}
where $j_{\mu a}^{L}$ and $j_{\mu a}^{NL}$ are the local and nonlocal
contributions, respectively. We can now check explicitly the current
conservation. With help of the equations of motion, we find 
\begin{eqnarray}
\partial _{\mu }j_{\mu a}^{L}\left( z\right) &=&\sum_{\omega ,\lambda
_{\omega }}\frac{m\langle \omega ,\lambda _{\omega }\left| z\rangle \langle
z|[\beta ,\lambda _{a}]\right| \omega ,\lambda _{\omega }\rangle }{\omega
-ie_{\lambda }\left( \omega \right) }-  \label{loc} \\
&&\sum_{\omega ,\lambda _{\omega }}\frac{\langle \omega ,\lambda _{\omega
}\left| z\rangle \langle z|\beta \lambda _{a}r\Phi r\right| \omega ,\lambda
_{\omega }\rangle -\langle \omega ,\lambda _{\omega }|\beta r\Phi r\lambda
_{a}|z\rangle \langle z|\omega ,\lambda _{\omega }\rangle }{\omega
-ie_{\lambda }\left( \omega \right) }  \nonumber \\
\partial _{\mu }j_{\mu a}^{NL}\left( z\right) &=&\sum_{\omega ,\lambda
_{\omega }}\frac{\langle \omega ,\lambda _{\omega }\left| z\rangle \langle
z|\beta \lambda _{a}r\Phi r\right| \omega ,\lambda _{\omega }\rangle
-\langle \omega ,\lambda _{\omega }|\beta r\Phi r\lambda _{a}|z\rangle
\langle z|\omega ,\lambda _{\omega }\rangle }{\omega -ie_{\lambda }\left(
\omega \right) }  \label{nonloc} \\
&&-\sum_{\omega ,\lambda _{\omega }}\frac{\langle \omega ,\lambda _{\omega
}\left| r|z\rangle \lambda _{a}\beta \Phi (z)\langle z|r\right| \omega
,\lambda _{\omega }\rangle -\langle \omega ,\lambda _{\omega }|r|z\rangle
\beta \Phi (z)\lambda _{a}\langle z|r|\omega ,\lambda _{\omega }\rangle }{%
\omega -ie_{\lambda }\left( \omega \right) }.  \nonumber
\end{eqnarray}
In the second equation we have used the identity 
\begin{eqnarray}
\frac{\partial }{\partial z_{\mu }}i\int \frac{d_{4}q}{(2\pi )^{4}}%
\int_{x}^{y}ds_{\mu }e^{i(z-s)\cdot q} &=&\int \frac{d_{4}q}{(2\pi )^{4}}%
\int_{x}^{y}d(s\cdot q)e^{i(z-s)\cdot q}=  \nonumber \\
i\int \frac{d_{4}q}{(2\pi )^{4}}(e^{i(z-y)\cdot q}-e^{i(z-x)\cdot q})
&=&i(\delta (z-y)-\delta (z-x)).
\end{eqnarray}
Combining the local and nonlocal pieces (\ref{loc},\ref{nonloc}) (which are
not separately conserved), and using the equations of motion for the $S$ and 
$P$ fields, Eqs. (\ref{eulag}), we find that 
\begin{equation}
\partial _{\mu }j_{\mu a}\left( z\right) =-\sum_{\omega ,\lambda _{\omega }}%
\frac{\langle \omega ,\lambda _{\omega }|z\rangle \lbrack \lambda _{a},\beta
m]\langle z|\omega ,\lambda _{\omega }\rangle }{\omega -ie_{\lambda }\left(
\omega \right) }.
\end{equation}
This immediately leads to the conservation of baryon and isospin current,
and, in the chiral limit of $m=0$, to the conservation of the axial current.
Note that the conservation laws are independent of the chosen path in the $P$%
-exponent.

Quantities involving space integrals of Noether currents (charges, $g_{A}$)
also do not depend on the path. Indeed, the integration over $z$ in Eq.
(\ref{jmua}) leads to 
\begin{equation}
\int \frac{d_{4}q}{(2\pi )^{4}}\delta (q)\int_{x}^{y}ds_{\mu }e^{-is\cdot
q}=\int_{x}^{y}ds_{\mu }=(y^{\mu }-x^{\mu }).  \label{bb}
\end{equation}
and subsequently, through the use of the identity 
\begin{equation}
i\left( y_{\mu }-x_{\mu }\right) \left\langle x\left| r\right|
y\right\rangle =\left\langle x\left| r_{\mu }\right| y\right\rangle ,
\label{ider1}
\end{equation}
to 
\begin{equation}
\int d_{3}z\,j_{\mu a}\left( z\right) =-\int \frac{d\omega }{2\pi i}%
\sum_{\lambda _{\omega }}\frac{\langle \lambda _{\omega }\left| (\beta
\lambda _{a}\gamma _{\mu }+r_{\mu }\lambda _{a}\beta \Phi r+\beta r\Phi
\lambda _{a}r_{\mu })\right| \lambda _{\omega }\rangle }{\omega -ie_{\lambda
}\left( \omega \right) }.  \label{ider11}
\end{equation}
Note that any reference to the choice of the path has disappeared. Thus,
charges, which are obtained from Eq. (\ref{ider11}) with $\mu =0$, or $%
g_{A}$, which has $\mu =3$ ({\em cf.} Sect. \ref{sec:ga}) are independent of
the path, hence are uniquely defined. One may easily generalize
this result to any Green's function in the soliton background for the case
where the external legs corresponding to Noether currents 
have vanishing four-momenta. The proof is
straightforward through the use of the identity (\ref{bb}).

\section{Straight-line paths}

\label{app:straight}

The expression (\ref{jmua}) is not, in general, suitable for calculating
observables, since the path is not specified. A popular choice of the path
is a straight line \cite{BowlerB,PlantB,bled99}: 
\begin{equation}
s_{\mu }=x_{\mu }+\alpha \left( y_{\mu }-x_{\mu }\right) \;\;\;\;\;ds_{\mu
}=d\alpha \left( y_{\mu }-x_{\mu }\right) .  \label{straight}
\end{equation}
Then in Eq. (\ref{jmua}) we have
\begin{eqnarray}
\int \frac{d_{4}q}{(2\pi )^{4}}\int_{x}^{y}ds_{\mu }e^{i(z-s)\cdot q}
&=&\int \frac{d_{4}q}{(2\pi )^{4}}\int_{0}^{1}d\alpha \left( y_{\mu }-x_{\mu
}\right) e^{i(z-x-\alpha \left( y-x\right) )\cdot q} \\
&=&\int_{0}^{1}d\alpha \left( y_{\mu }-x_{\mu }\right) \delta (z-x-\alpha
\left( y-x\right) ),
\end{eqnarray}
As a result, we find
\begin{eqnarray}
&&j_{\mu a}^{{\rm NL,straigh}}\left( z\right) =-\sum_{\omega ,\lambda
_{\omega }}\int d_{4}x\int d_{4}y\int_{0}^{1}d\alpha \,\delta (z-x-\alpha
\left( y-x\right) )\times \\
&&\frac{\langle \omega ,\lambda _{\omega }|x\rangle \langle x|r_{\mu
}|y\rangle \langle y|\lambda _{a}\beta \Phi r|\omega ,\lambda _{\omega
}\rangle +\langle \omega ,\lambda _{\omega }|r\beta \Phi \lambda
_{a}|x\rangle \langle x|r_{\mu }|y\rangle \langle y|\omega ,\lambda _{\omega
}\rangle }{i\omega +e_{\lambda }\left( \omega \right) }.  \nonumber
\label{jstr}
\end{eqnarray}
Since the fields $\Phi $ are stationary, we can rewrite Eq. (\ref{jmua}%
) in a simpler form 
\begin{eqnarray}
&&j_{\mu a}^{{\rm NL,straigh}}\left( \vec{z}\right) =-\int \frac{d\omega }{%
2\pi }\sum_{\lambda _{\omega }}\int d_{3}x\int d_{3}y\int_{0}^{1}d\alpha
\,\delta (\vec{z}-\vec{x}-\alpha \left( \vec{y}-\vec{x}\right) )\times \\
&&\frac{\langle \lambda _{\omega }|\vec{x}\rangle \langle \vec{x}|r_{\mu
}(\omega )|\vec{y}\rangle \langle \vec{y}|\lambda _{a}\beta \Phi r(\omega
)|\lambda _{\omega }\rangle +\langle \lambda _{\omega }|r(\omega )\beta \Phi
\lambda _{a}|\vec{x}\rangle \langle \vec{x}|r_{\mu }(\omega )|\vec{y}\rangle
\langle \vec{y}|\lambda _{\omega }\rangle }{i\omega +e_{\lambda }\left(
\omega \right) }.  \nonumber
\end{eqnarray}
In the second term above, we can interchange $x$ with $y$ and change the
integration variable $\alpha \rightarrow 1-\alpha $. This leads to a
manifestly Hermitian form, 
\begin{eqnarray}
j_{\mu a}^{{\rm NL,straigh}}\left( \vec{z}\right) &=&-\int \frac{d\omega }{%
2\pi }\sum_{\lambda _{\omega }}\int d_{3}x\int d_{3}y\int_{0}^{1}d\alpha
\,\delta (\vec{z}-\vec{x}-\alpha \left( \vec{y}-\vec{x}\right) )\times 
\nonumber \\
&&\frac{\langle \lambda _{\omega }|\vec{x}\rangle \langle \vec{x}|r_{\mu
}(\omega )|\vec{y}\rangle \langle \vec{y}|\lambda _{a}\beta \Phi r(\omega
)|\lambda _{\omega }\rangle }{i\omega +e_{\lambda }\left( \omega \right) }%
+h.c.  \label{form1}
\end{eqnarray}
which is used below.

\section{Weak-nonlocality approximation}

\label{app:leading}

For the case where the nonlocality scale $\Lambda $ is much larger than
other scales in the problem ({\em e.g.} the inverse soliton size) we can
commute the $r_{\mu }$ operator with $|x\rangle \langle x|$ \cite{ripka:book}
in Eq. (\ref{form1}), thereby obtaining 
\begin{eqnarray}
j_{\mu a}^{{\rm NL,weak}}\left( z\right) &=&-\sum_{\omega ,\lambda _{\omega
}}\int d_{4}x\int d_{4}y\int_{0}^{1}d\alpha \,\delta (z-x-\alpha \left(
y-x\right) )\times  \nonumber \\
&&\frac{\langle \omega ,\lambda _{\omega }|r_{\mu }|x\rangle \langle
x|y\rangle \langle y|\lambda _{a}\beta \Phi r|\omega ,\lambda _{\omega
}\rangle }{i\omega +e_{\lambda }\left( \omega \right) }+h.c.  \nonumber \\
&=&-\sum_{\omega ,\lambda _{\omega }}\frac{\langle \omega ,\lambda _{\omega
}|r_{\mu }|z\rangle \langle z|\lambda _{a}\beta \Phi r|\omega ,\lambda
_{\omega }\rangle }{i\omega +e_{\lambda }\left( \omega \right) }+h.c.
\end{eqnarray}
Finally, commuting $r_{\mu }$ and $|z\rangle \langle z|$ and using the
stationarity of $\Phi $ we find 
\begin{equation}
j_{\mu a}^{{\rm NL,weak}}\left( \vec{z}\right) =-\int \frac{d\omega }{2\pi }%
\sum_{\lambda _{\omega }}\frac{\langle \lambda _{\omega }|\vec{z}\rangle
\langle \vec{z}|r_{\mu }(\omega )\lambda _{a}\beta \Phi r(\omega )|\lambda
_{\omega }\rangle }{i\omega +e_{\lambda }\left( \omega \right) }+h.c.
\end{equation}
which is our current in the weak-nonlocality approximation. Note that some
arbitrariness is involved here. We could have placed the $|\vec{z}\rangle
\langle \vec{z}|$ operator anywhere between $\langle \lambda _{\omega }|$
and $|\lambda _{\omega }\rangle ,$ and that would lead to different
currents. However, all of these definitions become equal if the scale of the
nonlocality is much larger than other scales in the problem, {\em i.e.} in
the weak-nonlocality limit.

\section{Evaluation of observables}

\label{app:observables}

\bigskip We calculate observables both with the straight-line prescription
and in the weak-nonlocality limit. We work in the Minkowski space, by means of
the replacement 
\begin{equation}
\gamma _{0}\rightarrow i\beta ,\qquad j_{0a}(x)\rightarrow i\rho _{a}(x),
\end{equation}
with $\rho _{a}$ denoting the Minkowski charge density.

\subsection{Baryon mean square radius}

\label{sec:ms}

The baryon charge mean squared radius equals to 
\begin{equation}
\left\langle r^{2}\right\rangle _{B}=\int d_{3}z\,\,z^{2}\rho _{B}\left( 
\vec{z}\right) ,
\end{equation}
The contribution from the local part of the current is 
\begin{equation}
\left\langle r^{2}\right\rangle _{B}^{{\rm L}}=-\frac{1}{N_{c}}\int \frac{%
d\omega }{2\pi i}\sum_{\lambda _{\omega }}\int d_{3}z\frac{\langle \lambda
_{\omega }\left| \vec{z}\rangle z^{2}\langle \vec{z}\right| \lambda _{\omega
}\rangle }{\omega -ie_{\lambda }\left( \omega \right) }\,,
\end{equation}
while the nonlocal term gives

\begin{eqnarray}
\left\langle r^{2}\right\rangle _{B}^{{\rm NL,straight}} &=&-\frac{2}{iN_{c}}%
\int d_{3}z\,\,z^{2}\int \frac{d\omega }{2\pi i}\sum_{\lambda _{\omega
}}\int d_{3}x\int d_{3}y\int_{0}^{1}d\alpha \,\times  \nonumber \\
&&\delta (\vec{z}-\vec{x}-\alpha \left( \vec{y}-\vec{x}\right) )\frac{%
\langle \lambda _{\omega }|\vec{x}\rangle \langle \vec{x}|r_{0}(\omega )|%
\vec{y}\rangle \langle \vec{y}|\beta \Phi r(\omega )|\lambda _{\omega
}\rangle }{\omega -ie_{\lambda }\left( \omega \right) }.
\end{eqnarray}
We can now carry the $\alpha $ integration 
\begin{eqnarray}
\int d_{3}z\,z^{2}\int_{0}^{1}d\alpha \,\delta (\vec{z}-\vec{x}-\alpha
\left( \vec{y}-\vec{x}\right) ) &=&\int_{0}^{1}d\alpha \,(\vec{x}+\alpha
\left( \vec{y}-\vec{x}\right) )^{2}=  \nonumber \\
\vec{x}^{2}+\vec{x}\cdot (\vec{y}-\vec{x})+\frac{1}{3}\left( \vec{y}-\vec{x}%
\right) ^{2} &=&\frac{1}{2}\vec{x}^{2}-\frac{1}{6}\left( \vec{y}-\vec{x}%
\right) ^{2}+\frac{1}{2}\vec{y}^{2}
\end{eqnarray}
to obtain 
\begin{eqnarray}
\left\langle r^{2}\right\rangle _{B}^{{\rm NL,straight}} &=&\frac{1}{2}A+%
\frac{1}{6}B+\frac{1}{2}C, \\
A &=&-\frac{2}{iN_{c}}\int \frac{d\omega }{2\pi i}\sum_{\lambda _{\omega
}}\int d_{3}z\,z^{2}\frac{\langle \lambda _{\omega }|r_{0}(\omega )|\vec{x}%
\rangle \langle \vec{x}|\beta \Phi r(\omega )|\lambda _{\omega }\rangle }{%
\omega -ie_{\lambda }\left( \omega \right) },  \nonumber \\
B &=&-\frac{2}{iN_{c}}\int \frac{d\omega }{2\pi i}\sum_{\lambda _{\omega
}}\sum_{j=1}^{3}\frac{\langle \lambda _{\omega }|r_{0jj}(\omega )\beta \Phi
r(\omega )|\lambda _{\omega }\rangle }{\omega -ie_{\lambda }\left( \omega
\right) },  \nonumber \\
C &=&-\frac{2}{iN_{c}}\int \frac{d\omega }{2\pi i}\sum_{\lambda _{\omega
}}\int d_{3}z\,\,z^{2}\frac{\langle \lambda _{\omega }|\vec{x}\rangle
\langle \vec{x}|r_{0}(\omega )\beta \Phi r(\omega )|\lambda _{\omega
}\rangle }{\omega -ie_{\lambda }\left( \omega \right) }.  \nonumber
\end{eqnarray}
In the derivation of $B$ we have used the identity 
\begin{equation}
(\vec{x}-\vec{y})^{2}\langle \vec{x}|r_{0}(\omega )|\vec{y}\rangle
=-\sum_{j=1}^{3}\langle \vec{x}|r_{0jj}(\omega )|\vec{y}\rangle .
\end{equation}
The above formulas can be further decomposed into the valence and sea
contributions:

\begin{eqnarray}
\left\langle r^{2}\right\rangle _{B}^{{\rm L,val}} &=&{\rm z}_{{\rm val}%
}\int d_{3}z\,\,z^{2}|\langle \vec{z}|{\rm val}\rangle |^{2},  \nonumber \\
\left\langle r^{2}\right\rangle _{B}^{{\rm L,sea}} &=&-\frac{1}{N_{c}}\int 
\frac{d\omega }{2\pi }\sum_{\lambda _{\omega }}\frac{e_{\lambda }\left(
\omega \right) }{\omega ^{2}+e_{\lambda }^{2}\left( \omega \right) }\int
d_{3}z\,z^{2}|\langle \vec{z}|\lambda _{\omega }\rangle |^{2},  \nonumber \\
A^{{\rm val}} &=&4{\rm z}_{{\rm val}}e_{{\rm val}}\int d_{3}x\,x^{2}\langle 
{\rm val}|r^{\prime }(ie_{{\rm val}})|\vec{x}\rangle \langle \vec{x}|\beta
\Phi r(ie_{{\rm val}})|{\rm val}\rangle ,  \nonumber \\
A^{{\rm sea}} &=&\frac{4}{N_{c}}\int \frac{d\omega }{2\pi }\sum_{\lambda
_{\omega }}\frac{\omega ^{2}}{\omega ^{2}+e_{\lambda }^{2}\left( \omega
\right) }\int d_{3}x\,x^{2}\langle \lambda _{\omega }|r^{\prime }(\omega )|%
\vec{x}\rangle \langle \vec{x}|\beta \Phi r(\omega )|\lambda _{\omega
}\rangle ,  \nonumber \\
B^{{\rm val}} &=&16{\rm z}_{{\rm val}}e_{{\rm val}}\langle {\rm val}|(-\hat{%
\nabla}^{2})r^{\prime \prime \prime }(ie_{{\rm val}})\beta \Phi r(ie_{{\rm %
val}})|{\rm val}\rangle ,  \nonumber \\
B^{{\rm sea}} &=&\frac{16}{N_{c}}\int \frac{d\omega }{2\pi }\sum_{\lambda
_{\omega }}\frac{\omega ^{2}}{\omega ^{2}+e_{\lambda }^{2}\left( \omega
\right) }\langle \lambda _{\omega }|(-\hat{\nabla}^{2})r^{\prime \prime
\prime }(\omega )\beta \Phi r(\omega )|\lambda _{\omega }\rangle ,  \nonumber
\\
C^{{\rm val}} &=&4{\rm z}_{{\rm val}}e_{{\rm val}}\int d_{3}x\,x^{2}\langle 
{\rm val}|\vec{x}\rangle \langle \vec{x}|r^{\prime }(ie_{{\rm val}})\beta
\Phi r(ie_{{\rm val}})|{\rm val}\rangle ,  \nonumber \\
C^{{\rm sea}} &=&\frac{4}{N_{c}}\int \frac{d\omega }{2\pi }\sum_{\lambda
_{\omega }}\frac{\omega ^{2}}{\omega ^{2}+e_{\lambda }^{2}\left( \omega
\right) }\int d_{3}x\,x^{2}\langle \lambda _{\omega }|\vec{x}\rangle \langle 
\vec{x}|r^{\prime }(\omega )\beta \Phi r(\omega )|\lambda _{\omega }\rangle .
\end{eqnarray}
In the derivation we have used the notation $r^{\prime
}(k^{2})=d/dk^{2}\,r(k^{2}),${\em \ etc.}, and the fact that $r(\omega
)=r(-\omega )$.{\em \ }

In the weak-nonlocality approximation we obtain 
\begin{equation}
\left\langle r^{2}\right\rangle _{B}^{{\rm NL,weak}}=C.
\end{equation}

\subsection{$g_{A}$}

\label{app:ga}

We evaluate $g_{A}$ from the expression 
\begin{equation}
g_{A}=-\frac{2}{3}\int d_{3}z\,A_{33}(\vec{z}),
\end{equation}
where $-\frac{2}{3}$ is a factor coming from the cranking projection method 
\cite{CB86} and $A_{33}$ is the hedgehog matrix element of the third-space,
third-isospin component of the axial charge. According to Eq. (\ref
{ider11}) we find 
\begin{equation}
g_{A}=\frac{1}{3}\int \frac{d\omega }{2\pi i}\sum_{\lambda _{\omega }}\frac{%
\langle \lambda _{\omega }\left| (\beta \gamma _{3}\gamma _{5}\tau
_{3}+r_{3}\gamma _{5}\tau _{3}\beta \Phi r+\beta r\Phi \gamma _{5}\tau
_{3}r_{3})\right| \lambda _{\omega }\rangle }{\omega -ie_{\lambda }\left(
\omega \right) },
\end{equation}
which gives

\begin{eqnarray}
g_{A}^{{\rm L,val}} &=&-\frac{N_{c}}{3}{\rm z}_{{\rm val}}\langle {\rm val}%
\left| \sigma _{3}\tau _{3}\right| {\rm val}\rangle ,  \nonumber \\
g_{A}^{{\rm L,sea}} &=&\frac{1}{3}\int \frac{d\omega }{2\pi }\sum_{\lambda
_{\omega }}\frac{e_{\lambda }\left( \omega \right) \langle \lambda _{\omega
}\left| \sigma _{3}\tau _{3}\right| \lambda _{\omega }\rangle }{\omega
^{2}+e_{\lambda }^{2}\left( \omega \right) },  \nonumber \\
g_{A}^{{\rm NL,val}} &=&-\frac{4N_{c}}{3}{\rm z}_{{\rm val}}\langle {\rm val}%
\left| (-i\hat{\nabla}_{3})r^{\prime }\gamma _{5}\tau _{3}\beta \Phi
r\right| {\rm val}\rangle ,  \nonumber \\
g_{A}^{{\rm NL,sea}} &=&\frac{4}{3}\int \frac{d\omega }{2\pi }\sum_{\lambda
_{\omega }}\frac{e_{\lambda }\left( \omega \right) \langle \lambda _{\omega
}\left| (-i\hat{\nabla}_{3})r^{\prime }\gamma _{5}\tau _{3}\beta \Phi
r\right| \lambda _{\omega }\rangle }{\omega ^{2}+e_{\lambda }^{2}\left(
\omega \right) }.
\end{eqnarray}

\subsection{Isovector magnetic moment and mean square radius}

\label{sec:mu}

The isovector magnetic moment is obtained from the expression 
\begin{equation}
\mu _{I=1}=-\frac{1}{3}\varepsilon ^{3jm}\int d_{3}z\,\,z_{j}V_{3m}(\vec{z}),
\end{equation}
where the $-\frac{1}{3}$ is the cranking projection factor \cite{CB86,ANW83}
and $V_{3m}$ is the hedgehog matrix element of the $m$-space,
third-isospin component of the isovector current. For the local contribution
we find 
\begin{eqnarray}
\mu _{{\rm I=1}}^{{\rm L,val}} &=&-\frac{N_{c}}{6}{\rm z}_{{\rm val}}\int
d_{3}z\,\langle {\rm val}|\vec{z}\rangle \beta (\vec{z}\times \vec{\gamma}%
)_{3}\tau _{3}\langle \vec{z}|{\rm val}\rangle , \\
\mu _{{\rm I=1}}^{{\rm L,sea}} &=&\frac{1}{6}\int \frac{d\omega }{2\pi }%
\sum_{\lambda _{\omega }}\frac{e_{\lambda }\left( \omega \right) \int
d_{3}z\,\langle \lambda _{\omega }|\vec{z}\rangle \beta (\vec{z}\times \vec{%
\gamma})_{3}\tau _{3}\langle \vec{z}|\lambda _{\omega }\rangle }{\omega
^{2}+e_{\lambda }^{2}\left( \omega \right) },
\end{eqnarray}
For the nonlocal contribution with the straight-line prescription we have 
\begin{eqnarray}
\mu _{{\rm I=1}}^{{\rm NL,straight}} &=&\frac{\varepsilon ^{3jm}}{6}\int
d_{3}z\,z_{j}\int d_{3}x\int d_{3}y\int_{0}^{1}d\alpha \,\delta (\vec{z}-%
\vec{x}-\alpha \left( \vec{y}-\vec{x}\right) )\,R_{m},  \nonumber \\
R_{m} &=&\int \frac{d\omega }{2\pi i}\sum_{\lambda _{\omega }}\frac{\langle
\lambda _{\omega }|\vec{x}\rangle \langle \vec{x}|r_{m}(\omega )|\vec{y}%
\rangle \langle \vec{y}|\tau _{3}\beta \Phi r(\omega )|\lambda _{\omega
}\rangle }{\omega -ie_{\lambda }\left( \omega \right) }+h.c.
\end{eqnarray}
Then 
\begin{eqnarray}
\mu _{{\rm I=1}}^{{\rm NL,straight}} &=&\frac{\varepsilon ^{3jm}\,}{6}\int
d_{3}x\int d_{3}y\int_{0}^{1}d\alpha (\vec{x}+\alpha \left( \vec{y}-\vec{x}%
\right) )_{j}\,\,R_{m}  \nonumber \\
&=&\frac{\varepsilon ^{3jm}}{6}\,\int d_{3}x\int d_{3}y(x_{j}+\frac{1}{2}%
\left( y_{j}-x_{j}\right) )\,\,R_{m}.
\end{eqnarray}
Through the use of the identities $\,\left( y_{j}-x_{j}\right) \langle \vec{x%
}|r_{m}(\omega )|\vec{y}\rangle =-i\langle \vec{x}|r_{mj}(\omega )|\vec{y}%
\rangle $ and $\varepsilon ^{3jm}r_{mj}(\omega )=0$ we obtain 
\begin{eqnarray}
\mu _{{\rm I=1}}^{{\rm NL,straight}} &=&\frac{\varepsilon ^{3jm}}{6}\,\int
d_{3}x\,\,\int \frac{d\omega }{2\pi i}\sum_{\lambda _{\omega }}\frac{\langle
\lambda _{\omega }|\vec{x}\rangle x_{j}\langle \vec{x}|r_{m}(\omega )\tau
_{3}\beta \Phi r(\omega )|\lambda _{\omega }\rangle }{\omega -ie_{\lambda
}\left( \omega \right) }+h.c.  \nonumber \\
&&
\end{eqnarray}
Since $\varepsilon ^{3jm}x_{j}r_{m}(\omega )=\varepsilon ^{3jm}x_{j}(-2i\hat{%
\nabla}_{m})r^{\prime }(\omega )=2L_{3}r^{\prime }(\omega )$, with $\vec{L}$
being the orbital angular momentum operator, we can write 
\begin{equation}
\mu _{{\rm I=1}}^{{\rm NL,straight}}=\frac{1}{3}\int \frac{d\omega }{2\pi i}%
\sum_{\lambda _{\omega }}\frac{\langle \lambda _{\omega }|L_{3}r_{m}(\omega
)\tau _{3}\beta \Phi r(\omega )|\lambda _{\omega }\rangle }{\omega
-ie_{\lambda }\left( \omega \right) }+h.c.
\end{equation}
Now we notice that the same expression follows when we use the
weak-nonlocality approximation, therefore 
\begin{equation}
\mu _{I=1}^{NL,straight}=\mu _{I=1}^{{\rm NL,weak}}
\end{equation}
Finally, 
\begin{eqnarray}
\mu _{{\rm I=1}}^{{\rm NL,val}} &=&-\frac{2N_{c}}{3}{\rm z}_{{\rm val}%
}\,\langle {\rm val}|r^{\prime }(ie_{{\rm val}})L_{3}\tau _{3}\beta \Phi |r|%
{\rm val}\rangle , \\
\mu _{{\rm I=1}}^{{\rm NL,sea}} &=&\frac{2}{3}\int \frac{d\omega }{2\pi }%
\sum_{\lambda _{\omega }}\frac{e_{\lambda }\left( \omega \right) \langle
\lambda _{\omega }|r^{\prime }(\omega )L_{3}\tau _{3}\beta \Phi r|\lambda
_{\omega }\rangle }{\omega ^{2}+e_{\lambda }^{2}\left( \omega \right) }.
\end{eqnarray}
Since $[r^{\prime },L_{3}]=0$, we are free to interchange the order of $%
r^{\prime }$ and $L_{3}$ in the above formulas.

The isovector magnetic mean square radius is defined as 
\begin{equation}
\left\langle r^{2}\right\rangle _{{\rm m,I=1}}=-\frac{1}{3\mu _{I=1}}%
\varepsilon ^{3jm}\int d_{3}z\,\,z_{j}z^{2}V_{3m}(\vec{z}),
\end{equation}
For the local part we find 
\begin{eqnarray}
\left\langle r^{2}\right\rangle _{{\rm m,I=1}}^{{\rm L,val}} &=&-\frac{N_{c}%
}{6\mu _{I=1}}{\rm z}_{{\rm val}}\int d_{3}z\,z^{2}\langle {\rm val}|\vec{z}%
\rangle \beta (\vec{z}\times \vec{\gamma})_{3}\tau _{3}\langle \vec{z}|{\rm %
val}\rangle , \\
\left\langle r^{2}\right\rangle _{{\rm m,I=1}}^{{\rm L,sea}} &=&\frac{1}{%
6\mu _{I=1}}\int \frac{d\omega }{2\pi }\sum_{\lambda _{\omega }}\frac{%
e_{\lambda }\left( \omega \right) \int d_{3}z\,z^{2}\langle \lambda _{\omega
}|\vec{z}\rangle \beta (\vec{z}\times \vec{\gamma})_{3}\tau _{3}\langle \vec{%
z}|\lambda _{\omega }\rangle }{\omega ^{2}+e_{\lambda }^{2}\left( \omega
\right) },
\end{eqnarray}
For the nonlocal part we have 
\begin{eqnarray}
\left\langle r^{2}\right\rangle _{{\rm m,I=1}}^{{\rm NL,straight}} &=&\frac{%
\varepsilon ^{3jm}}{6\mu _{I=1}}\int d_{3}z\,z^{2}z_{j}\int d_{3}x\int
d_{3}y\int_{0}^{1}d\alpha \,\delta (\vec{z}-\vec{x}-\alpha \left( \vec{y}-%
\vec{x}\right) )\,R_{m}  \nonumber \\
&=&\frac{\varepsilon ^{3jm}}{6\mu _{I=1}}\,\int_{0}^{1}d\alpha (\vec{x}%
+\alpha \left( \vec{y}-\vec{x}\right) )^{2}(\vec{x}+\alpha \left( \vec{y}-%
\vec{x}\right) )_{j}\,\,R_{m} \\
&=&\frac{\varepsilon ^{3jm}}{6\mu _{I=1}}\int d_{3}x\int d_{3}y\left[ \frac{1%
}{2}x^{2}x_{j}+\frac{1}{2}y^{2}x_{j}-\frac{1}{6}\left( \vec{y}-\vec{x}%
\right) ^{2}x_{j}\right] R_{m}.  \nonumber
\end{eqnarray}
Finally 
\begin{eqnarray}
\left\langle r^{2}\right\rangle _{{\rm m,I=1}}^{{\rm NL,straight}} &=&\frac{1%
}{2}A_{m}+\frac{1}{6}B_{m}+\frac{1}{2}C_{m}, \\
A_{m}^{{\rm val}} &=&-\frac{2N_{c}}{3\mu _{I=1}}{\rm z}_{{\rm val}}\int
d_{3}z\,z^{2}\langle {\rm val}|r^{\prime }(ie_{{\rm val}})L_{3}|\vec{z}%
\rangle \tau _{3}\beta \Phi (\vec{z})\langle \vec{z}|r|{\rm val}\rangle , 
\nonumber \\
A_{m}^{{\rm sea}} &=&\frac{2}{3\mu _{I=1}}\int \frac{d\omega }{2\pi }%
\sum_{\lambda _{\omega }}\frac{e_{\lambda }\left( \omega \right) \int
d_{3}z\,z^{2}\langle \lambda _{\omega }|r^{\prime }(\omega )L_{3}|\vec{z}%
\rangle \tau _{3}\beta \Phi (\vec{z})\langle \vec{z}|r|\lambda _{\omega
}\rangle }{\omega ^{2}+e_{\lambda }^{2}\left( \omega \right) },  \nonumber \\
B_{m}^{{\rm val}} &=&-\frac{8N_{c}}{3\mu _{I=1}}{\rm z}_{{\rm val}}\langle 
{\rm val}|(-\hat{\nabla}^{2})r^{\prime \prime \prime }(ie_{{\rm val}%
})L_{3}\tau _{3}\beta \Phi r|{\rm val}\rangle ,  \nonumber \\
B_{m}^{{\rm sea}} &=&\frac{8}{3\mu _{I=1}}\int \frac{d\omega }{2\pi }%
\sum_{\lambda _{\omega }}\frac{e_{\lambda }\left( \omega \right) \langle
\lambda _{\omega }|(-\hat{\nabla}^{2})r^{\prime \prime \prime }(\omega
)L_{3}\tau _{3}\beta \Phi r|\lambda _{\omega }\rangle }{\omega
^{2}+e_{\lambda }^{2}\left( \omega \right) },  \nonumber \\
C_{m}^{{\rm val}} &=&-\frac{2N_{c}}{3\mu _{I=1}}{\rm z}_{{\rm val}}\int
d_{3}z\,z^{2}\langle {\rm val}|r^{\prime }(ie_{{\rm val}})|\vec{z}\rangle
\langle \vec{z}|L_{3}\tau _{3}\beta \Phi r|{\rm val}\rangle ,  \nonumber \\
C_{m}^{{\rm sea}} &=&\frac{2}{3\mu _{I=1}}\int \frac{d\omega }{2\pi }%
\sum_{\lambda _{\omega }}\frac{e_{\lambda }\left( \omega \right) \int
d_{3}z\,z^{2}\langle \lambda _{\omega }|r^{\prime }(\omega )|\vec{z}\rangle
\langle \vec{z}|L_{3}\tau _{3}\beta \Phi r|\lambda _{\omega }\rangle }{%
\omega ^{2}+e_{\lambda }^{2}\left( \omega \right) }.  \nonumber
\end{eqnarray}
In the weak-nonlocality limit
\begin{equation}
\left\langle r^{2}\right\rangle _{{\rm m,I=1}}^{{\rm NL,weak}}=C_{m}.
\end{equation}


\end{document}